\documentclass[preprint,3p,12pt]{elsarticle}

\usepackage{amssymb}
\usepackage{amsmath}
\usepackage{hyperref}
\hypersetup{
    colorlinks=true,
    linkcolor=blue,
    filecolor=blue,      
    urlcolor=black,
    citecolor=blue,
        }

\journal{Journal of Computational Physics}

\begin{document}

\begin{frontmatter}



\title{A numerical framework for phoretic particles}


\author[label1,label2]{Zhe Gou}
\author[label2]{Alexander Farutin}
\author[label2]{Chaouqi Misbah\corref{cor1}}
%\ead{}
\cortext[cor1]{chaouqi.misbah@univ-grenoble-alpes.fr}
\affiliation[label1]{organization={College of Mechanical and Electrical Engineering, Central South University},
            city={Changsha},
            postcode={410083}, 
            country={China}}
\address[label2]{Universit\'e Grenoble Alpes, CNRS, LIPhy, 
            Grenoble,
            F-38000, 
            France}

\begin{abstract}
We develop a numerical a framework to study phoretic particle dynamics in two dimensions.
The particles are modeled as chemically active rigid circles, which can emit or absorb a solute into surrounding fluid.
The interaction between particles and solute induces a slip flow on particle surfaces, and the solute is advected by the fluid flow and diffuses with a constant diffusivity.
The fluid-structure interaction is resolved by a boundary integral method accelerated by Ewald-like decomposition.
The sharp resolution of moving boundaries for solute kinetics is performed thanks to  an overlapping mesh method.
The framework is validated separately for the Stokes problem and the advection--diffusion problem, reaching relatively high order of accuracy.
Moreover, we employ the framework to more general problems, including particles in nearly infinite domain and straight channels, and multiparticle motions.
%The spatial convergence shows a second order accuracy.

\end{abstract}




\end{frontmatter}


\section{Introduction}
\label{sec:intro}
In the past decades, phoretic particles (including rigid particles and droplets) have drawn increasing attention in experimental, theoretical, and numerical studies \cite{Michelin13,Izri14,Herminghaus14,Michelin14,Sondak16,
Jin17,Thutupalli18,Hu19,Farutin22,Li22,Michelin23}.
This problem involves an intimate coupling of fluid flow and solute transport with a moving boundary.
More precisely, the fluid flow is induced by solute concentration gradient on particle surface, while the solute, besides diffusion and reactions, is advected by the flow \cite{Michelin13,Izri14,Michelin14}.
In addition, the flow and solute concentration are affected by freely swimming particles and fixed boundaries \cite{Herminghaus14,Jin17,Jin19,Izzet20}.
The interaction among multiple particles might lead to collective behaviors {such as traveling lines, dynamic crystallites \cite{Thutupalli18}, caging \cite{Hokmabad22a}, and rotating clusters \cite{Hokmabad22b}}.
In order to study these complex dynamics, a general numerical framework is needed. This constitutes the main motivation of the present work.

Numerical modeling of phoretic particles is not a trivial task, due to the nonlinear character of the  solute advection by the fluid flows as well as to the presence of moving boundary conditions.
Most numerical frameworks are designed for specific problems, which either deal with simple geometries or simplified models.
For simple geometric configurations like the case of a  single particle in an unconfined domain or the case of  two-particle interactions, the full coupling of flow and solute field can be resolved using polar coordinates \cite{Michelin13,Michelin14,Sondak16,Hu19,Li22} or bi-polar coordinates \cite{Lippera20,Desai21}.
However, this does not apply to the case of  more particles or to the case of a  confined geometry, like in the presence of bounding walls (e.g. straight channels).
To investigate collective behaviors for multiple particles, simplified models have been used. A simplification consisted of  considering the particles  as point sources \cite{Lippera21}, instead of being of finite size. Another simplification considered the case  of rapid diffusion and rapid viscous transport \cite{Mont15,Singh19,Rojas21}.
In these simplified models, the hydrochemical coupling is discarded, and the solute transport reduces to either unsteady diffusion or Laplace problems.
For this reason, some behaviors which result from hydrodynamic interactions, like particle attraction \cite{Jin19,Hokmabad22b}, cannot be reproduced.
{ Recently, some efforts have been directed towards  accounting for more general geometries, larger numbers of particles, and nonlinearity of solute advection.}
By combining the embedded boundary method (EBM) with adaptive mesh refinement, the self-propulsion of an isotropic active particle in cylindrical pipes have been studied \cite{Picella22}. 
Others studies used the immersed boundary method (IBM) for both flow and solute field \cite{Chen21,Yang23}, which transfers sharp boundary problems into diffuse boundary problems.

A full numerical framework for phoretic particles consists of two parts: the Stokes flow and the advection-diffusion problem for solute, which both contain moving boundaries.
As particle motions are determined by solute distributions on their surfaces \cite{Michelin13}, the key point is to resolve boundary conditions on these moving boundaries.
For the Stokes flow, the moving boundary problem can be solved by Green's function, also know as boundary integral method (BIM) \cite{Pozrikidis92}. 
However, most boundary integral based methods are not appropriate for the present study due to following reasons.
First, as the solute is advected by the flow, the boundary integral equation should be calculated on all sampling points in the fluid domain, not just on boundaries.
Second, with the increase of particle numbers, the computational complexity scales as $O(N^2)$, where $N$ is proportional to the product of the number of particles and the number of sampling points on each particle.
To overcome these limitations, several accelerated techniques have been developed, such as the particle-particle-particle-mesh (P$^3$M) method \cite{Deserno98}, the particle-mesh-Ewald (PME) method \cite{Essmann95,Guckel99}, and general geometry Ewald-like method (GGEM) \cite{Hernandez07,Kumar12}.

There are other numerical techniques for simulating fluid flow and advection-diffusion problems with moving boundaries.
One candidate is the EBM or the cut-cell method (CCM), by which the Cartesian meshes are cut and reconstructed by moving boundaries \cite{McCorquodale01,Schneiders16}.
The reconstructed meshes are body-fitted in the vicinity of boundaries, which provide sharp resolution of boundaries and interfaces, while the remainder remains fixed Cartesian meshes.
The overlapping mesh method (OMM) employs body-fitted meshes to several overlapping subdomains, and interpolates values on subdomain interfaces \cite{Chesshire90,Henshaw08}.
Moreover, it allows these decomposed subdomains to move independently, suitable for moving boundary problems \cite{Prewitt00,Henshaw06,Merrill19}.
Another popular technique is the immersed boundary method (IBM) \cite{Goldstein93,Peskin02,Wang20}.
Boundary conditions are considered as source terms, and are spreaded to neighboring mesh points with a smoothed delta function, creating a smeared interface with finite thickness.
To obtain sharp interfaces, the immersed interface method (IIM) applies interpolation or extrapolation to derive approximate boundary values or normal derivatives on neighboring mesh points \cite{Lai01,Hu18}.
Similar ideas are applied to lattice Boltzmann method (LBM) as modified bounce-back boundary conditions \cite{Lallemand03,Huang16,Zhang19}.
Last but not least, particle-based methods are also used for studies of moving boundary problems, such as multiparticle collision dynamics (MPC) \cite{Noguchi04,Noguchi05}, dissipative particle dynamics (DPD) \cite{Fedosov10,Gao17,Xu23}, and smoothed particle hydrodynamics (SPH) \cite{Tanaka05,Hosseini09}, to name but a few.

In preliminary studies, we first attempted to modify an IBM-based framework \cite{Zhang18,Gou21} to model phoretic particles numerically.
The Stokes flow was solved using the combined immersed boundary-lattice Boltzmann method (IBLBM) \cite{Shen17}.
To account for low inertial effect, the Reynolds number was set to 0.05.
The LBM with modified bounce-back boundary condition was employed for the advection and diffusion of solute \cite{Zhang19}.
Although this IBM-based framework could reproduce the symmetry breaking from stationary to straight motion for a circular phoretic particle, we found that the critical P\'eclet number $Pe$ for straight motion was smaller (within 5 to 10 $\%$) than the analytical finding \cite{Hu19}.
Here $Pe$ is the ratio of advection and diffusion.
Moreover, we could  obtain only qualitative agreement for meandering, circular, and chaotic motions.
 %, including phoretic velocities and critical $Pe$ for transition between motions.
Further analysis showed that the IBM and the modified bounce-back boundary condition were only 1st-order accurate on the particle surface.
%Hence we conclude that this IBM-based framework might be used to explore particle dynamics qualitatively, but not accurate enough for quantitative studies.
In fact, quantitative disagreements have been observed in other studies \cite{Chen21,Yang23} when using IBM-based methods.
For a circular phoretic particle, the critical $Pe$ for straight motion was also smaller than the benchmark in Ref. \cite{Yang23}, and meandering or circular motions were not reported in that work.
For a spherical phoretic particle, chaotic motion was reported for $Pe > 15$ in Ref. \cite{Chen21}, while a more accurate result should be $Pe \geq 24.2$ \cite{Lin20,Hu22}.
These works indicate that a better alternative for the IBM should be proposed for quantitative studies of phoretic particles.

In this work, we develop a fast boundary integral method (FBIM) for arbitrary geometries, inspired by the GGEM \cite{Hernandez07}.
The FBIM combines the advantages of BIM and IBM, being sufficiently efficient and accurate at the same time.
For the advection-diffusion problem, we employ an OMM with second order accuracy in space.
The subdomains attached to particles can move freely in the background fluid domain, resolving moving boundary conditions.
The rest of this paper is organized as follows.
In Section \ref{sec:problem}, we first recall the governing equations for the motion of phoretic particles.
Then we present the numerical implementation of the FBIM and the OMM in Sections \ref{sec:FBIM} and \ref{sec:OMM}, respectively.
The solution procedure is presented in Section \ref{sec:procedure}.
{In Section \ref{sec:validation}, the FBIM and the OMM are validated separately, and convergence studies are performed.
Lastly, the combined FBIM-OMM is employed to solve several general problems in Section \ref{sec:applications}, and the effects of confined geometries and multiparticle interactions are explored.}

\section{Problem formulation}\label{sec:problem}
We consider a two-dimensional (2D) domain $\mit\Omega$ of circular shaped particles of radius $a$ immersed in a Newtonian fluid of viscosity $\mu$.
These particles are chemically active and can emit or absorb a solute into the surrounding fluid with an isotropic flux $\mathcal{A}$.
The solute diffuses with a molecular diffusivity $\mathcal{D}$ and is advected by the fluid flow.
There is a chemical consumption/production proportional to the local solute concentration with coefficient $\mathcal{B}$.
The interaction of the solute molecules with the particles induces a slip tangential flow on the particle surfaces with a mobility $\mathcal{M}$ \cite{Michelin14}.

In the following, we adopt the non-dimensionalization of Ref. \cite{Michelin13} where the length, fluid velocity, the solute concentration, and the pressure {(or stress)} are scaled by the characteristic values $a$, $|\mathcal{AM}|/\mathcal{D}$, $a|\mathcal{A}|/\mathcal{D}$, and $\mu|\mathcal{AM}|/a\mathcal{D}$, respectively.
The governing equations in dimensionless form are
\begin{equation}\label{eq:stokes}
\nabla p(\boldsymbol{x}, t) = \nabla^2 \boldsymbol{u}(\boldsymbol{x}, t), 
\nabla \cdot \boldsymbol{u}(\boldsymbol{x}, t) = 0,
\end{equation}
\begin{equation}\label{eq:ad}
\frac{\partial c(\boldsymbol{x}, t)}{\partial t} + \boldsymbol{u}(\boldsymbol{x}, t) \cdot \nabla c(\boldsymbol{x}, t) = \frac{1}{Pe} \nabla^2 c(\boldsymbol{x}, t) - \beta c(\boldsymbol{x}, t),
\end{equation}
where $p$, $\boldsymbol{u}$ and $c$ are the dimensionless fluid pressure, velocity and solute concentration, $Pe = a|\mathcal{AM}|/\mathcal{D}^2$ is the P\'eclet number which describes the ratio of advection and diffusion over  the transport of solute, and $\beta = a\mathcal{BD}/|\mathcal{AM}|$ is the dimensionless consumption/production coefficient.

At the surface of a particle $\mit\Gamma_p$ or a fixed boundary $\mit\Gamma_f$, the boundary conditions for the flow and the concentration field read
\begin{equation}\label{eq:ub}
\boldsymbol{u}(\boldsymbol{x}, t)|_{\mit\Gamma_p} = M \nabla_s c(\boldsymbol{x}, t)|_{\mit\Gamma_p} + \boldsymbol{U}_p(t) + \boldsymbol{\mit\Omega}_p(t) \times (\boldsymbol{x} - \boldsymbol{X}_p),
\boldsymbol{u}(\boldsymbol{x}, t)|_{\mit\Gamma_f} = \boldsymbol{0},
\end{equation}
\begin{equation}\label{eq:cb}
\boldsymbol{n} \cdot \nabla c(\boldsymbol{x}, t)|_{\mit\Gamma_p} = - A,
\boldsymbol{n} \cdot \nabla c(\boldsymbol{x}, t)|_{\mit\Gamma_f} = 0,
\end{equation}
where $\boldsymbol{U}_p$ and $\boldsymbol{\mit\Omega}_p$ are the translational and rotational velocities of the particle, $\boldsymbol{X}_p$ is the center of the particle, $\nabla_s$ is the surface gradient operator, $\boldsymbol{n}$ is the unit outward normal, $M = \mathcal{M}/|\mathcal{M}|$ is the dimensionless mobility and $A = \mathcal{A}/|\mathcal{A}|$ is the dimensionless activity.
In order to trigger self-propulsion in infinite fluid, $A$ and $M$ must have the same sign \cite{Michelin13,Hu19}.
%For definiteness we consider in the following that $A = M = 1$.
The particle moves as a rigid body, and its center $\boldsymbol{X}_p$ and orientation angle $\mit\Theta_p$ are given by
\begin{equation}\label{eq:XpThetap}
\frac{d\boldsymbol{X}_p}{dt} = \boldsymbol{U}_p(t),
\frac{d\mit\Theta_p}{dt} = \boldsymbol{\mit\Omega}_p(t).
\end{equation}
In the Stokes regime, there are two other constraints on each particle in the absence of external force: the force-free and torque-free conditions
\begin{equation}\label{eq:force_free}
\int_{\mit\Gamma_p} \boldsymbol{\sigma}(\boldsymbol{x}, t) \cdot \boldsymbol{n} dS(\boldsymbol{x}) = \boldsymbol{0},
\int_{\mit\Gamma_p} (\boldsymbol{x} - \boldsymbol{X}_p) \times \left[ \boldsymbol{\sigma}(\boldsymbol{x}, t) \cdot \boldsymbol{n} \right] dS(\boldsymbol{x}) = 0,
\end{equation}
where $\boldsymbol{\sigma} = -p \boldsymbol{I} + \nabla \boldsymbol{u} + \nabla \boldsymbol{u}^T$ is the flow stress tensor, {and $S$ is the curvilinear coordinate on the particle boundary $\mit\Gamma_p$}.

\section{Fast boundary integral method for fluid}\label{sec:FBIM}
The implementation of the FBIM is given as follows.
We first recall the boundary integral equation, and then derive the analytical solution due to a point force {based on Ewald-like decomposition}.
Later we apply the {decomposed solution} to calculate the single layer integral on boundaries.
At last, a feedback forcing method is introduced to derive surface forces applied by rigid body on the fluid.

\subsection{Boundary integral equation}

The flow field is described by the Stokes equation (\ref{eq:stokes}) with Dirichlet boundary conditions (\ref{eq:ub}).
A general solution can be obtained by the boundary integral equation \cite{Pozrikidis92}
\begin{equation}\label{eq:bim}
u_i(\boldsymbol{x}) = \frac{1}{4 \pi} \sum_{n=1}^{N_b} \int_{\mit\Gamma_n} G_{ij}(\boldsymbol{x}, \boldsymbol{X}) F_j(\boldsymbol{X}) dS(\boldsymbol{X}),
\end{equation}
where $\mit\Gamma_n$ is the contour of the $n$th boundary, $\boldsymbol{X}$ is a point on this boundary, $\boldsymbol{G}$ is the 2D Green's function, $\boldsymbol{F}$ is the surface force applied by the boundary on the fluid, and the summations are over all the $N_b$ boundaries.
For an infinite domain, the free-space Green's function \cite{Pozrikidis92} is used as
\begin{equation}\label{eq:green}
G_{ij}(\boldsymbol{x}, \boldsymbol{x}') = -\delta_{ij} \ln(|\boldsymbol{x} - \boldsymbol{x}'|) + \frac{(x_i - x'_i) (x_j - x'_j)}{|\boldsymbol{x} - \boldsymbol{x}'|^2}.
\end{equation}
The FBIM is derived based on Eqs. (\ref{eq:bim}) and (\ref{eq:green}).

%\begin{table}[hbt!]
%\caption{Model parameters for simulation set up}
%\label{tab_parameters_simu}
%\centering

%\begin{tabular}{l c c}
%\hline
%Notation & Value & Physical meaning \\
%\hline
%$R_0$ & 3 $\mu$m & Characteristic radius of RBC \\
%$\tau$ & 0.65 & Reduced area of RBC \\
%$k_b$ & $3 \times 10^{-19}$ J & Membrane bending modulus \\
%$\eta_{out}$ & 0.001 Pa$\cdot$s & Plasma viscosity \\
%$\lambda$ & 6 & Viscosity ratio between plasma and cytosol \\
%$D_{ATP}$ & $2.36 \times 10^{-10}$ m$^2$/s & Diffusion coefficient of ATP in plasma \\
%\hline
%\end{tabular}

%\end{table}

\subsection{Decomposed solution due to a point force}
\label{sec:point_force}

\begin{figure}[hbt!]
\centering
	\includegraphics[width=0.8\textwidth]{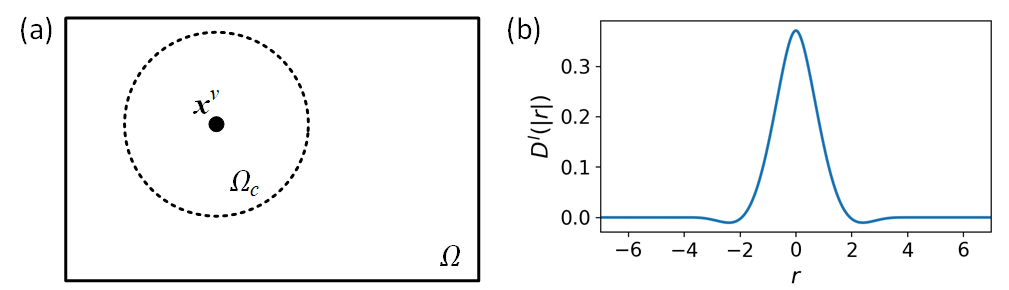}
	\caption{(a) Schematic of the point force at $\boldsymbol{x}^v$ and the near neighbor region $\mit\Omega_c$.
	(b) Long-range density $D^l$ [see Eq. (\ref{eq:Dl})] as the function of $r$.
	Here the cutoff distance is set as $r_c = 4$.}
	\label{fig:fbim_schematic}
\end{figure}

We start from the simplest case, which is the flow due to a point force:
\begin{equation}\label{eq:point_force}
\nabla p(\boldsymbol{x}) = \nabla^2 \boldsymbol{u}(\boldsymbol{x}) + \boldsymbol{g}^v\delta(|\boldsymbol{x} - \boldsymbol{x}^v|), 
\nabla \cdot \boldsymbol{u}(\boldsymbol{x}) = 0,
\end{equation}
where $\boldsymbol{g}^v$ is the force applied at a point $\boldsymbol{x}^v$ (see Fig. \ref{fig:fbim_schematic}a).
The resulted velocity is
\begin{equation}
u_i(\boldsymbol{x}) = \frac{1}{4 \pi} G_{ij}(\boldsymbol{x}, \boldsymbol{x}^v) g_j^v.
\end{equation}
{Inspired by other Ewald-summation-based methods \cite{Essmann95,Guckel99,Hernandez07},} the idea of the FBIM is to split the Dirac-delta density $\delta(r)$ in Eq. (\ref{eq:point_force}) into {a sum of a short-range density $D^s(r)$ containing the singularity, and a sufficiently smooth long-range density $D^l(r) = \delta(r) - D^s(r)$.}
%smoothly distributed long-range density $D^l(r)$ and another short-range density $D^s(r)$.
Here $r$ refers to the magnitude of the vector from any position to the pole of singularity, $\boldsymbol{r} = \boldsymbol{x} - \boldsymbol{x}^v$.
{The short-range density is chosen in such a way that for any point beyond a certain cutoff distance $r_c$ (i.e. $r > r_c$), the flow velocity due to the short-range force density $\boldsymbol{g}^v D^s(r)$ is zero.}
%For any point beyond a certain cutoff distance $r_c$ (i.e. $r > r_c$), the flow velocity due to the short-range force density $\boldsymbol{g}^v D^s(r)$ is zero.
%The long-range density is given by $D^l(r) = \delta(r) - D^s(r)$}.
%It should be noted that the total density is still the Dirac-delta function, that is $D^l(r) + D^s(r) = \delta(\boldsymbol{r})$.

The solution of the Stokes and continuity equation is also decomposed into two parts.
For the long-range part, the solution $\boldsymbol{u}^l(\boldsymbol{x})$ and $p^l(\boldsymbol{x})$ due to the long-range force density satisfies the following set of equations
\begin{equation}\label{eq:long}
\nabla p^l(\boldsymbol{x}) = \nabla^2 \boldsymbol{u}^l(\boldsymbol{x}) + \boldsymbol{g}^v D^l(|\boldsymbol{x} - \boldsymbol{x}^v|), 
\nabla \cdot \boldsymbol{u}^l(\boldsymbol{x}) = 0.
\end{equation}
This equation can be solved by any Stokes solver, and we will use a Fourier spectral method in this study.
The short-range solution $\boldsymbol{u}^s(\boldsymbol{x})$ and $p^s(\boldsymbol{x})$ satisfies the short-range set of equations
\begin{equation}\label{eq:short}
\nabla p^s(\boldsymbol{x}) = \nabla^2 \boldsymbol{u}^s(\boldsymbol{x}) + \boldsymbol{g}^v D^s(|\boldsymbol{x} - \boldsymbol{x}^v|), 
\nabla \cdot \boldsymbol{u}^s(\boldsymbol{x}) = 0.
\end{equation}
This equation will be solved by using a short-range Green's function for any points within the cutoff distance ($r \leq r_c$).
Once the two problems are solved, the overall solution is obtained as
\begin{equation}
\boldsymbol{u}(\boldsymbol{x}) = \boldsymbol{u}^l(\boldsymbol{x}) + \boldsymbol{u}^s(\boldsymbol{x}),
p(\boldsymbol{x}) = p^l(\boldsymbol{x}) + p^s(\boldsymbol{x}).
\end{equation}

\subsubsection{Long-range solution}\label{sec:long_range}
To solve the long-range problem, a proper long-range density $D^l(r)$ should be chosen. 
The first constraint on $D^l(r)$ is the finite cutoff distance $r_c$, such that
\begin{equation}\label{eq:Dl_cutoff}
D^l(r) = 0, \forall r \geq r_c.
\end{equation}
Despite this, in order to have a smooth solution of $\boldsymbol{u}^l(\boldsymbol{x})$, its derivatives should be continuous at $r = r_c$ as
\begin{equation}\label{eq:Dl_smooth}
\left. \frac{\partial^k D^l}{\partial r^k} \right|_{r = r_c} = 0, k = 1, 2, 3, ... , k_{max}
\end{equation}
where {$k_{max}$ is the cut-off order of the derivatives.}

The second constraint is related to the solution induced by the point force.
As follows from the above, the solution driven by the long-range force density is the same as that driven by the point force for any point beyond the cutoff distance.
Define $\mit\Omega_c$ as the near neighbor region around the singular point $\boldsymbol{x}^v$ within the cutoff distance $r_c$ (see Fig. \ref{fig:fbim_schematic}a), then the following equation is valid for any point $\boldsymbol{x} \in \mit\Omega \setminus \mit\Omega_c$:
\begin{equation}\label{eq:long_range_green}
u_i^l(\boldsymbol{x}) = \frac{1}{4 \pi} G_{ij}(\boldsymbol{x}, \boldsymbol{x}^v) g_j ^v = \frac{1}{4 \pi} \int_{\mit\Omega_c} G_{ij}(\boldsymbol{x}, \boldsymbol{x}') g_j^v D^l(|\boldsymbol{x}' - \boldsymbol{x}^v|) d\boldsymbol{x}'.
\end{equation}

We found a 6th-order polynomial which satisfies the above constraints as (see Fig. \ref{fig:fbim_schematic}b)
\begin{equation}\label{eq:Dl}
D^l(r) = \left\{
\begin{aligned}
&0, & r \geq r_c \\
&\frac{56}{3 \pi r_c^2} \left[-\frac{25}{2}(\frac{r}{r_c})^6 + 54(\frac{r}{r_c})^5 - 90(\frac{r}{r_c})^4 + 70(\frac{r}{r_c})^3 - \frac{45}{2}(\frac{r}{r_c})^2 + 1 \right], & r < r_c
\end{aligned}
\right.
.
\end{equation}
The detailed derivation is given in \ref{sec:appendix_a}.
Actually, it turns out that  an expansion of   $D^l(r)$ as higher-order polynomial with only even degree terms leads to higher order accuracy.
It should be noted that the choice of long-range density is not unique.
Any functions that satisfy Eqs. (\ref{eq:Dl_cutoff})--(\ref{eq:long_range_green}) would be valid.
Moreover, the long-range density $D^l$ is strictly equal to 0 for any point beyond the cutoff distance.
This is different from the screening function in GGEM \cite{Hernandez07}, which is  quasi-Gaussian and decays exponentially to 0.

To solve Eq. (\ref{eq:long}), the discrete Fourier series approximation is employed in the periodic $x$ and $y$ directions.
The two components of the long-range velocity $\boldsymbol{u}^l = (u_x^l, u_y^l)$ is expressed as
\begin{equation}
u_x^l(\boldsymbol{x}) + i u_y^l(\boldsymbol{x}) = \sum_{m = -N_x/2}^{N_x/2} \sum_{n = -N_y/2}^{N_y/2} \hat{u}_{mn}^l e^{i 2\pi mx/L_x} e^{i 2\pi ny/L_y},
\end{equation}
where $N_x$ and $N_y$ are the number of modes in the corresponding series approximation, and $\hat{u}_{mn}^l$ is the complex amplitude of the harmonic.
Similarly, the long-range force density $\boldsymbol{f}^l(\boldsymbol{x}) = \boldsymbol{g}^v D^l(|\boldsymbol{x} - \boldsymbol{x}^v|)$ is also represented by discrete Fourier series as
\begin{equation}
f_x^l(\boldsymbol{x}) + i f_y^l(\boldsymbol{x}) = \sum_{m = -N_x/2}^{N_x/2} \sum_{n = -N_y/2}^{N_y/2} \hat{f}_{mn}^l e^{i 2\pi mx/L_x} e^{i 2\pi ny/L_y},
\end{equation}
where $\hat{f}_{mn}^l$ is the complex amplitude of the harmonic. 
The computational domain is a rectangle, and is discretized by means of equally spaced Cartesian mesh.
The number of mesh points in $x$ and $y$ directions are set equal to the number of Fourier modes $N_x$ and $N_y$ respectively, so that the fast Fourier transform (FFT) can be used to transfer data between real and Fourier space.
The FFT and reverse FFT are calculated by the cuFFT library \cite{cuFFT22}, which is accelerated by GPU.
In practice, Eq. (\ref{eq:long}) is first transformed into Fourier space using FFT.
Then the harmonic's amplitude $\hat{u}_{mn}^l$ is calculated as
\begin{equation}
\hat{u}^l_{mn} = \frac{1}{2(m^2 + n^2)} \left(\hat{f}^l_{mn} - \frac{m^2 + i2 mn - n^2}{m^2 + n^2}\hat{f}^{l*}_{-m-n} \right), mn \neq 0,
\end{equation}
where $\hat{f}^{l*}_{-m-n}$ is the complex conjugate of $\hat{f}^{l}_{-m-n}$.
The 0th mode $\hat{u}^l_{00}$ 
%should be treated separately, as it
corresponds to  the average velocity in the computational domain, which is zero, unless there is an imposed flow (not considered here).
%In general, any value satisfies periodic boundary conditions.
% The Stokes equation ensures a unique solution for  $\hat{u}^l_{00}$.
%In the present work, we chose $\hat{u}^l_{00} = 0$ for all cases, so that the average velocity is 0 along both $x$ and $y$ directions.
%For other cases like confined Poiseuille flow, one can set a proper $\hat{u}^l_{00}$ value to satisfy corresponding boundary conditions.}
Finally, a reverse FFT is employed to compute the long-range velocity $\boldsymbol{u}^l$ in real space.
For any point not lying on the mesh, the velocity is obtained by bi-cubic interpolation with fourth order accuracy in space.

\subsubsection{Short-range solution}
The short-range density is 
\begin{equation}\label{eq:Ds}
D^s(r) = \delta (\boldsymbol{r}) - D^l(r).
\end{equation}
The short-range velocity is then obtained as
\begin{equation}\label{eq:us}
u_i^s(\boldsymbol{x}) = \frac{1}{4 \pi} \int_{\mit\Omega_c} G_{ij}(\boldsymbol{x}, \boldsymbol{x}') g_j^v D^s(|\boldsymbol{x}' - \boldsymbol{x}^v|) d\boldsymbol{x}'.
\end{equation}
According to Eqs. (\ref{eq:Dl}) and (\ref{eq:Ds}), $u_i^s(\boldsymbol{x}) = 0$ for any point $\boldsymbol{x} \in \mit\Omega \setminus \mit\Omega_c$.
Hence Eq. (\ref{eq:us}) is solved for the points within the cutoff distance (i.e. $\boldsymbol{x} \in \mit\Omega_c$) only.
We define the short-range Green's function as
\begin{equation}
G_{ij}^s (\boldsymbol{x}, \boldsymbol{x}^v) = \int_{\mit\Omega_c} G_{ij}(\boldsymbol{x}, \boldsymbol{x}') D^s(|\boldsymbol{x}' - \boldsymbol{x}^v|) d\boldsymbol{x}'.
\end{equation}
Then the short-range velocity reads
\begin{equation}
u_i^s(\boldsymbol{x}) = \frac{1}{4 \pi} g_j^v G_{ij}^s(\boldsymbol{x}, \boldsymbol{x}^v).
\end{equation}
The derivation of the short-range Green's function is given in \ref{sec:appendix_b}.

\subsection{Single layer integral}
The velocity due to a boundary $\mit\Gamma$ is given by the single layer integral according to Eq. (\ref{eq:bim}).
Applying the same decomposition as Sec. \ref{sec:point_force}, we rewrite this equation in the following form
\begin{equation}
\begin{aligned}
u_i(\boldsymbol{x}) &= \frac{1}{4 \pi} \int_{\mit\Gamma} G_{ij}(\boldsymbol{x}, \boldsymbol{X}) \left[ \int_{\mit\Omega} F_j(\boldsymbol{X}) \delta (\boldsymbol{x}' - \boldsymbol{X}) d\boldsymbol{x}' \right] dS(\boldsymbol{X})\\
 &= \frac{1}{4 \pi} \int_{\mit\Gamma} \left\{ \int_{\mit\Omega} G_{ij}(\boldsymbol{x}, \boldsymbol{x}') F_j(\boldsymbol{X}) \left[ D^l (|\boldsymbol{x}' - \boldsymbol{X}|) + D^s (|\boldsymbol{x}' - \boldsymbol{X}|) \right] d\boldsymbol{x}' \right\} dS(\boldsymbol{X}).
\end{aligned}
\end{equation}
The contribution due to the long-range density is
\begin{equation}
u_i^l(\boldsymbol{x}) = \frac{1}{4 \pi} \int_{\mit\Omega} G_{ij}(\boldsymbol{x}, \boldsymbol{x}') \left[ \int_{\mit\Gamma} F_j(\boldsymbol{X}) D^l (|\boldsymbol{x}' - \boldsymbol{X}|) dS(\boldsymbol{X}) \right] d\boldsymbol{x}',
\end{equation}
and the long-range force density is
\begin{equation}\label{eq:long_force}
\boldsymbol{f}^l(\boldsymbol{x}) = \int_{\mit\Gamma} \boldsymbol{F}(\boldsymbol{X}) D^l (|\boldsymbol{x} - \boldsymbol{X}|) dS(\boldsymbol{X}).
\end{equation}
Next, the contribution due to the short-range density is
\begin{equation}\label{eq:short_bim}
\begin{aligned}
u_i^s(\boldsymbol{x}) &= \frac{1}{4 \pi} \int_{\mit\Gamma} F_j(\boldsymbol{X}) \left[ \int_{\mit\Omega} G_{ij}(\boldsymbol{x}, \boldsymbol{x}') D^s (|\boldsymbol{x}' - \boldsymbol{X}|) d\boldsymbol{x}' \right] dS(\boldsymbol{X})\\
 &= \frac{1}{4 \pi} \int_{\mit\Gamma} F_j(\boldsymbol{X}) G_{ij}^s(\boldsymbol{x}, \boldsymbol{X}) dS(\boldsymbol{X}).
\end{aligned}
\end{equation}
%The full solution is obtained by solving the long-range problem represented by Eq. (\ref{eq:long_ibm}) and short-range problem represented by Eq. (\ref{eq:short_bim}).

Here we describe the numerical implementation of the above problems.
The surface $\mit\Gamma$ is discretized into $N_e$ elements.
For the long-range problem, we first need to know the long-range force density $\boldsymbol{f}^l$ by computing the integral in Eq. (\ref{eq:long_force}).
Its discretized form is written as
\begin{equation}
\boldsymbol{f}^l (\boldsymbol{x}) = \sum_{k = 1}^{N_e} \int_{\mit\Gamma_k} \boldsymbol{F}(\boldsymbol{X}) D^l (|\boldsymbol{x} - \boldsymbol{X}|) dS(\boldsymbol{X}),
\end{equation}
where $\mit\Gamma_k$ denotes the $k$th element.
The above integral can be evaluated by a simple trapezoidal rule.
Then the long-range velocity $\boldsymbol{u}^l$ is solved for as described in Sec. \ref{sec:point_force}.
For the shor-range problem, the boundary integral in Eq. (\ref{eq:short_bim}) is discretized as
\begin{equation}\label{eq:short_range_integral_element}
\boldsymbol{u}^s (\boldsymbol{x}) = \frac{1}{4 \pi} \sum_{\mit\Gamma_k \in \mit\Omega_c} \int_{\mit\Gamma_k} \boldsymbol{F}(\boldsymbol{X}) \boldsymbol{G}^s (\boldsymbol{x}, \boldsymbol{X}) dS(\boldsymbol{X}).
\end{equation}
As the integrand in Eq. (\ref{eq:short_range_integral_element}) includes logarithmic singularity at boundaries $\boldsymbol{x} = \boldsymbol{X}$, a special quadrature rule \cite{Crow93} is used to ensure sufficient accuracy.
It should be noted that the summation in Eq. (\ref{eq:short_range_integral_element}) is calculated over elements within the near neighbor region of $\boldsymbol{x}$, i.e. $\mit\Gamma_k \in \mit\Omega_c$.
The contribution of other elements has been included in the long-range solution.

\subsection{Rigid body dynamics}
In the present study, particles and geometric boundaries are considered as rigid bodies.
The surface force $\boldsymbol{F}$ on these boundaries can be calculated by considering them as elastic bodies with a large elastic modulus.
An alternative method is to apply a feedback force \cite{Goldstein93} as
\begin{equation}\label{eq:feedback_force}
\boldsymbol{F}(\boldsymbol{X}, t) = \int_0^t \alpha \left[ \boldsymbol{V}(\boldsymbol{X}, t') - \boldsymbol{u}(\boldsymbol{X}, t') \right] dt',
\end{equation}
so that the desired velocity $\boldsymbol{V}$ is enforced on the boundary.
Here $\alpha$ is a positive coefficient.
In numerical application, the time integral in Eq. (\ref{eq:feedback_force}) is computed in an iterative form as
\begin{equation}\label{eq:feedback_force}
\boldsymbol{F}(\boldsymbol{X}, t + \Delta t) = \boldsymbol{F}(\boldsymbol{X}, t) + \alpha \left[ \boldsymbol{V}(\boldsymbol{X}, t) - \boldsymbol{u}(\boldsymbol{X}, t) \right] \Delta t,
\end{equation}
where $\Delta t$ is the time step.

The desired velocity depends on the boundary conditions (\ref{eq:ub}).
For a fixed boundary with no-slip boundary condition, $\boldsymbol{V} = \boldsymbol{0}$.
For a freely-moving circular phoretic particle, the desired velocity is the summation of the slip velocity, rigid body translation, and rotation
\begin{equation}\label{eq:V}
\boldsymbol{V}(\boldsymbol{X}) = M \nabla_s c(\boldsymbol{X}) + \boldsymbol{U}_p + \boldsymbol{\mit\Omega}_p \times (\boldsymbol{X} - \boldsymbol{X}_p).
\end{equation}
As the surface force $\boldsymbol{F}$ equals to the hydrodynamic traction $\boldsymbol{\sigma} \cdot \boldsymbol{n}$ on boundaries, the force-free and torque-free conditions in Eq. (\ref{eq:force_free}) read
\begin{equation}\label{eq:force_torque}
\int_{\mit\Gamma_p} \boldsymbol{F}(\boldsymbol{X}) dS(\boldsymbol{X}) = \boldsymbol{0},
\int_{\mit\Gamma_p} (\boldsymbol{X} - \boldsymbol{X}_p) \times \boldsymbol{F}(\boldsymbol{X}) dS(\boldsymbol{X}) = 0.
\end{equation}
The 4 variables $\boldsymbol{F}$, $\boldsymbol{V}$, $\boldsymbol{U}_p$, and $\boldsymbol{\mit\Omega}_p$ can be obtained by solving the system of Eqs. (\ref{eq:feedback_force}), (\ref{eq:V}) and (\ref{eq:force_torque}).
Then the motion of each particle is integrated in time by a forward Euler scheme as
\begin{equation}\label{eq:X}
\boldsymbol{X}_p(t + \Delta t) = \boldsymbol{X}_p(t) + \Delta t \boldsymbol{U}_p(t),
\mit\Theta_p(t + \Delta t) = \mit\Theta_p(t) + \Delta t \boldsymbol{\mit\Omega}_p(t).
\end{equation}

For circular particles, pre-conditioning is applied for better convergence.
In a polar coordinate system $(1, \theta)$ attached to the particle center $\boldsymbol{X}_p$, the surface force $\boldsymbol{F}[\boldsymbol{X}(\theta)]$, the surface velocity $\boldsymbol{u}[\boldsymbol{X}(\theta)]$, and the surface concentration $c[\boldsymbol{X}(\theta)]$ are all periodic, and can be expressed by Fourier series as
\begin{equation}\label{eq:force_fourier}
F_x(\theta, t) + i F_y(\theta, t) = \sum_{k = - \infty}^{\infty} \hat{F}_k(t) e^{ik\theta},
\end{equation}
\begin{equation}
u_x(\theta, t) + i u_y(\theta, t) = \sum_{k = - \infty}^{\infty} \hat{u}_k(t) e^{ik\theta},
\end{equation}
\begin{equation}
c(\theta, t) = \sum_{k = - \infty}^{\infty} \hat{c}_k(t) e^{ik\theta},
\end{equation}
where $\hat{F}_k$, $\hat{u}_k$, and $\hat{c}_k$ are the complex amplitudes.
Applying force-free, torque-free, and velocity divergence-free conditions, one finds
\begin{equation}
\hat{F}_0 = 0, \text{Im}(\hat{F}_1) = 0, \text{Re}(\hat{u}_1) = 0,
\end{equation}
where Re and Im denote real and imaginary parts, respectively.
The value of Re$(\hat{F}_1)$ does not affect the flow field, and is set to 0.
From time $t$ to $t + \Delta t$, a small change of surface force is applied, inducing a change of surface velocity as
\begin{equation}
\hat{\alpha}[\boldsymbol{V}(\theta, t) - \boldsymbol{u}(\theta, t)] = \int_0^{2\pi} \boldsymbol{G}(\theta, \phi) \left[\boldsymbol{F}(\phi, t + \Delta t) - \boldsymbol{F}(\phi, t)\right] d\phi.
\end{equation}
Here $\hat{\alpha}$ is a coefficient between 0 to 1.
This equation can be rewritten using Fourier series as
\begin{equation}
U_{p_x}(t) + i U_{p_y}(t) + i |\boldsymbol{\mit\Omega}_p(t)| e^{i\theta} + \sum_{k = -\infty}^{\infty} \left[M(1-k) \hat{c}_{k-1} (t) - \hat{u}_k (t)\right] e^{ik\theta} = \sum_{k \neq 0, 1} \frac{\hat{F}_k (t + \Delta t) - \hat{F}_k (t)}{4\hat{\alpha}|k|} e^{ik\theta}.
\end{equation}
Solving the equation for the $k$th Fourier harmonics, we obtain
\begin{equation}
\boldsymbol{U}_p(t) = \begin{bmatrix}
\text{Re}[\hat{u}_0(t)] - M \text{Re}[\hat{c}_1(t)] \\
\text{Im}[\hat{u}_0(t)] + M \text{Im}[\hat{c}_1(t)]
\end{bmatrix},
\end{equation}
\begin{equation}
|\boldsymbol{\mit\Omega}_p(t)| = \text{Im}[\hat{u}_1(t)],
\end{equation}
\begin{equation}
\hat{F}_k(t + \Delta t) = \hat{F}_k(t) + 4\hat{\alpha} |k| \left[M (1 - k) \hat{c}_{k-1}(t) - \hat{u}_k(t)\right], k \neq 0, 1.
\end{equation}
The surface force $\boldsymbol{F}(\theta, t + \Delta t)$ can then be calculated using Fourier series.

To ensure long-term stability of simulations, we perform a small correction of particle position when a particle approaches other particles or fixed boundaries.
The correction is introduced thanks to the Morse potential \cite{Liu04}
\begin{equation}
\Phi(d) = D_e[e^{2\gamma(d_0 - d)} - 2e^{\gamma(d_0 - d)}],
\end{equation}
where $D_e$ is the surface energy, $\gamma$ is the scaling factor, $d$ is the distance between two boundaries, and $d_0$ is the zero force distance, set to $3\Delta x$ for all simulations in the present work.
Consider the $m$th and the $n$th particles located at $\boldsymbol{X}_{p_m}$ and $\boldsymbol{X}_{p_n}$.
The distance between the two particle surfaces is $d_{mn} = r_{mn} - 2$, where $\boldsymbol{r}_{mn} = \boldsymbol{X}_{p_m} - \boldsymbol{X}_{p_n}$.
The correction for the $m$th particle due to the $n$th particle is
\begin{equation}\label{eq:repulsion}
\boldsymbol{X}^c_{mn} = \left\{
\begin{aligned}
& 0, & d_{mn} > d_0 \\
& -\Delta t \frac{\partial \Phi(d_{mn})}{\partial d_{mn}} \frac{\boldsymbol{r}_{mn}}{r_{mn}}, & d_{mn} \leq d_0 \\
\end{aligned}
\right.
.
\end{equation}
For the $m$th particle close to the $n$th fixed boundary, the distance $d_{mn}$ is the minimum distance between the particle surface and the fixed boundary, and $\boldsymbol{r}_{mn}$ is the normal vector of the fixed boundary pointing into the particle center.
The correction is then calculated using Eq. (\ref{eq:repulsion}).
The position of the $m$th particle after correction is
\begin{equation}
\boldsymbol{X}_{p_m} (t + \Delta t) = \boldsymbol{X}_{p_m} (t) + \Delta t \boldsymbol{U}_{p_m} (t) + \sum_{m \neq n}{\boldsymbol{X}^c_{mn} (t)}.
\end{equation}

\section{Overlapping mesh method for solute}\label{sec:OMM}

The detailed implementation of the OMM is given in this section.
First we decompose the computational domain into two overlapping subdomains.
Then the advection-diffusion problem is discretized in each subdomain.
Last but not least, the treatment of moving boundary is introduced.

\subsection{Domain decomposition}

\begin{figure}[hbt!]
\centering
	\includegraphics[width=1\textwidth]{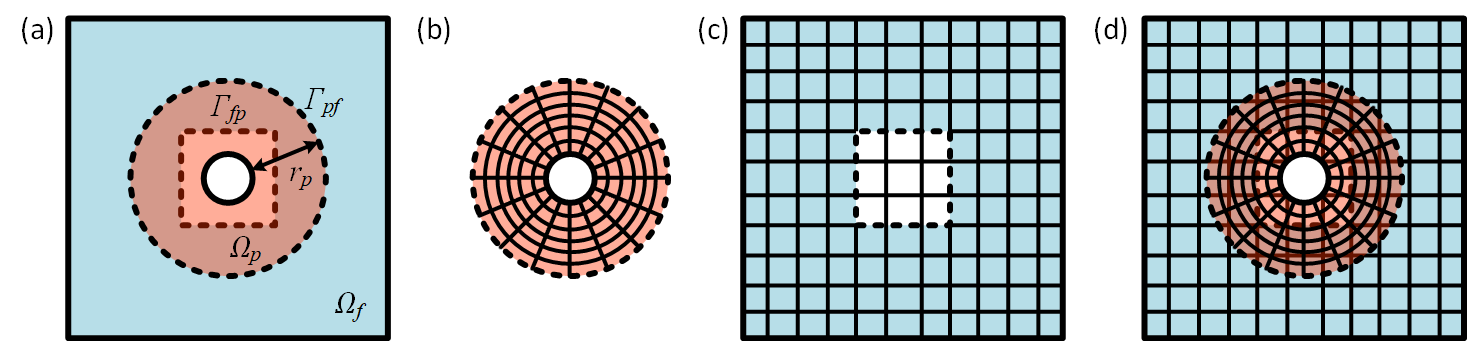}
	\caption{Schematics of the OMM.
	(a) Domain decomposition.
	(b) Polar mesh in the particle subdomain $\mit\Omega_p$
	(c) Cartesian mesh in the fluid subdomain $\mit\Omega_f$
	(d) Overlapped polar mesh and Cartesian mesh.
	For an annular subdomain, the difference of radius between the inner and outer circles is defined as the subdomain size $r_p$.
	The solid lines represent global boundaries, and the dashed lines represent interface boundaries.}
	\label{fig:omm_schematic}
\end{figure}

{The main idea of the OMM is to divide a computational domain into several overlapping subdomains \cite{Chesshire90}.}
For the present work, we consider a circular particle immersed in a rectangular domain.
The global solution domain $\mit\Omega$ is decomposed into two overlapping subdomains, the fluid subdomain $\mit\Omega_f$ and the particle subdomain $\mit\Omega_p$, so that $\mit\Omega = \mit\Omega_f \cup \mit\Omega_p$, as shown in Fig. \ref{fig:omm_schematic}a.
{For an annular subdomain like $\mit\Omega_p$, the difference of radius between the inner and outer circles is defined as the subdomain size $r_p$.}
Boundary conditions (\ref{eq:cb}) are employed on global boundaries (solid lines in Fig. \ref{fig:omm_schematic}a).
{Solute concentration }at interface boundaries $\mit\Gamma_{fp}$ and $\mit\Gamma_{pf}$ (dashed lines in Fig. \ref{fig:omm_schematic}a) is equal to contiguous value in {adjacent} subdomains, leading to additional Dirichlet boundary conditions {[as shown in Eqs. (\ref{eq:adfbi}) and (\ref{eq:adpbi})]}.
As the particle swims, the subdomain $\mit\Omega_p$ moves with the particle, so do the interface boundaries.
Here we use polar coordinates $(r, \theta)$ in $\mit\Omega_p$, where the origin is set at the particle center $\boldsymbol{X}_p$.
The mapping between two subdomains is written as
\begin{equation}
\boldsymbol{r} = \boldsymbol{x} - \boldsymbol{X}_p = r \begin{bmatrix}
\cos(\theta + \mit\Theta_p) \\
\sin(\theta + \mit\Theta_p) 
\end{bmatrix},
\end{equation}
where $\mit\Theta_p$ is the orientation angle of the particle.
The governing equations with boundary conditions in each subdomain read
\begin{equation}\label{eq:adf}
\frac{\partial c(\boldsymbol{x}, t)}{\partial t} + \boldsymbol{u}(\boldsymbol{x}, t) \cdot \nabla c(\boldsymbol{x}, t) = \frac{1}{Pe} \nabla^2 c(\boldsymbol{x}, t) - \beta c(\boldsymbol{x}, t),
\boldsymbol{x} \in \mit\Omega_f,
\end{equation}
\begin{equation}\label{eq:adfb}
\boldsymbol{n} \cdot \nabla c(\boldsymbol{x}, t)|_{\mit\Gamma_f} = 0,
\end{equation}
\begin{equation}\label{eq:adfbi}
c(\boldsymbol{x}, t)|_{\mit\Gamma_{fp}} = c(\boldsymbol{r}, t)|_{\mit\Gamma_{fp} \cap \mit\Omega_p},
\end{equation}
and
\begin{equation}\label{eq:adp}
\frac{\partial c(\boldsymbol{r}, t)}{\partial t} + \boldsymbol{u}_p(\boldsymbol{r}, t) \cdot \nabla c(\boldsymbol{r}, t) = \frac{1}{Pe} \nabla^2 c(\boldsymbol{r}, t) - \beta c(\boldsymbol{r}, t),
\boldsymbol{r} \in \mit\Omega_p,
\end{equation}
\begin{equation}\label{eq:adpb}
\boldsymbol{n} \cdot \nabla c(\boldsymbol{r}, t)|_{\mit\Gamma_p} = - A,
\end{equation}
\begin{equation}\label{eq:adpbi}
c(\boldsymbol{r}, t)|_{\mit\Gamma_{pf}} = c(\boldsymbol{x}, t)|_{\mit\Gamma_{pf} \cap \mit\Omega_f},
\end{equation}
where $\boldsymbol{u}_p = \boldsymbol{u} - \boldsymbol{U}_p - \boldsymbol{\mit\Omega}_p \times \boldsymbol{r}$ is the relative velocity in the particle subdomain $\mit\Omega_p$.

\subsection{Spatial discretization}
\begin{figure}[hbt!]
\centering
	\includegraphics[width=0.5\textwidth]{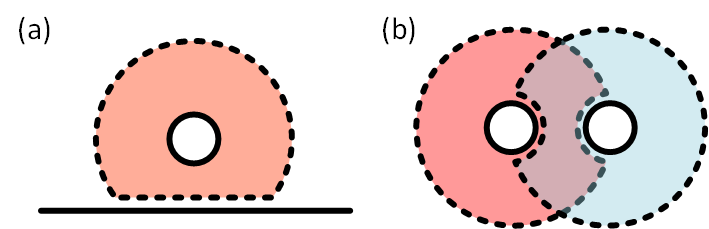}
	\caption{Schematics of interface boundaries.
	(a) A particle close to a global boundary.
	(b) Two particles close to each other.
	The solid lines represent global boundaries, and the dashed lines represent interface boundaries.}
	\label{fig:interface_schematic}
\end{figure}

The two subdomains are discretized separately.
The particle subdomain $\mit\Omega_p$ is discretized with polar mesh, and the outermost mesh points from the interface boundary $\mit\Gamma_{pf}$ (see Fig. \ref{fig:omm_schematic}b).
The fluid subdomain $\mit\Omega_f$ is discretized with Cartesian mesh, and its interface boundary $\mit\Gamma_{fp}$ is set as the Cartesian mesh points around the particle (see Fig. \ref{fig:omm_schematic}c).
A schematic of overlapping meshes is shown in Fig. \ref{fig:omm_schematic}d.
As the particle swims in the fluid, the polar mesh translates and rotates with the particle, while the Cartesian mesh remains fixed.
{It should be noted that the particle interface boundaries may evolve} due to particle movement, as shown in Fig. \ref{fig:interface_schematic}. 

\begin{figure}[hbt!]
\centering
	\includegraphics[width=0.6\textwidth]{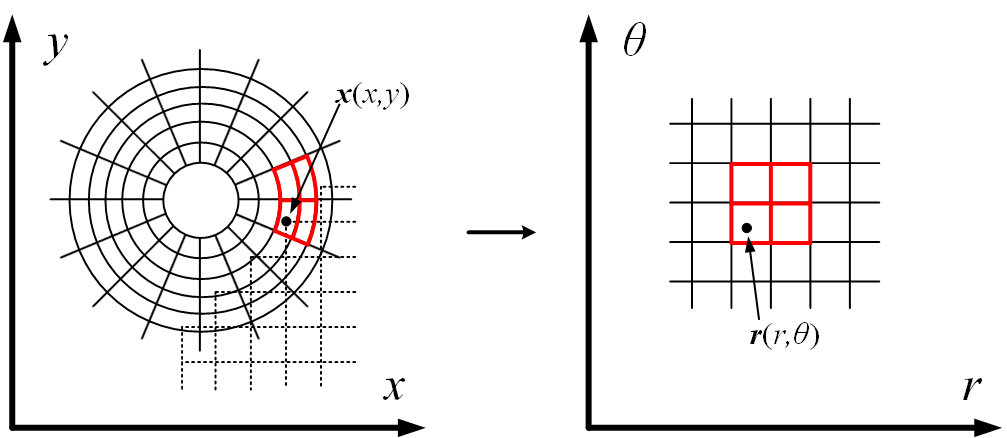}
	\caption{Interpolation of interface boundary conditions.
	The point $\boldsymbol{x}(x, y)$ in Cartesian coordinates is mapped to $\boldsymbol{r}(r, \theta)$ in polar coordinates.}
	\label{fig:interpolation_schematic}
\end{figure}

In each subdomain, the finite difference method is employed to solve the advection-diffusion problem numerically. 
The second-order centered difference is used to approximate the first and second derivatives in Eqs. (\ref{eq:adf}) and (\ref{eq:adp}).
For the Neumann boundary conditions (\ref{eq:adfb}) and (\ref{eq:adpb}), we apply the second-order forward or backward difference approximation.
The interface boundary conditions (\ref{eq:adfbi}) and (\ref{eq:adpbi}) are obtained by interpolation from the other subdomain.
Consider the situation shown in Fig. \ref{fig:interpolation_schematic} as an example, in which the concentration at point $\boldsymbol{x}(x,y) \in \mit\Gamma_{fp}$ is going to be interpolated from that in the particle subdomain $\mit\Omega_p$.
First, the point $\boldsymbol{x}$ is mapped from Cartesian coordinates to polar coordinates as $\boldsymbol{r}(r, \theta)$, then the interpolation is calculated in a rectangular grid as
\begin{equation}
c(\boldsymbol{r}) = \sum_{i=i_0}^{i_r} \sum_{j=j_0}^{j_r} \gamma_{ij} c_{ij},
\end{equation}
where $\gamma_{ij}$ is the interpolation weight, $c_{ij}$ is the solute concentration at grid points, and $(i_0, j_0)$ and $(i_r, j_r)$ are the lower left corner and the upper right corner of the interpolation stencil, respectively.
To maintain second-order accuracy of the global solution, a bi-quadratic interpolation stencil is used \cite{Chesshire90} (see the red grids in Fig. \ref{fig:interpolation_schematic}).

\subsection{Moving boundary treatment}
\begin{figure}[hbt!]
\centering
	\includegraphics[width=0.5\textwidth]{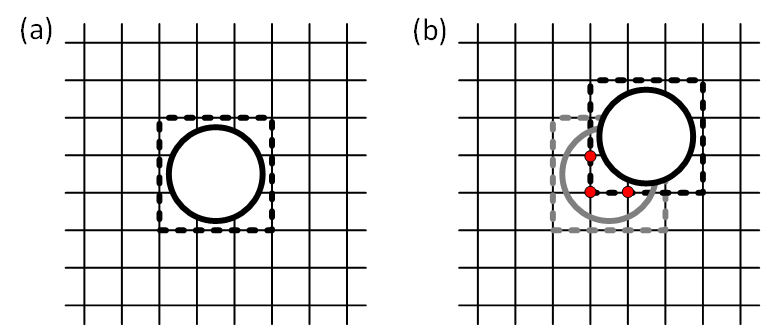}
	\caption{Refilling of mesh points (red dots) at two consecutive time steps.
	The solid lines indicate the particle, and the dashed lines indicate the interface boundaries $\mit\Gamma_{fp}$ of the fluid subdomain $\mit\Omega_f$.}
	\label{fig:moving_boundary_schematic}
\end{figure}

In the Cartesian mesh, some mesh points are covered by the particle, as shown in Fig. \ref{fig:moving_boundary_schematic}a.
At these points, concentration is set as 0.
As the particle moves, these mesh points may become uncovered, and solute concentrations should be evaluated at these points (see the red dots in Fig. \ref{fig:moving_boundary_schematic}b).
Similar situations happen to polar mesh points when the particle get close to fixed boundaries in the fluid subdomain.
To update concentrations at these newly appeared points, we employ interpolations from other subdomains using Eqs. (\ref{eq:adfbi}) and (\ref{eq:adpbi}) after advancing the particle.

Another problem arises when there are multiple particles.
As two particles get too close to each other, there might not be enough Cartesian mesh points for interpolation using Eq. (\ref{eq:adpbi}).
At this time, the interface boundary conditions of one particle are interpolated from the subdomain of the other particle.
This ensures accuracy without local mesh refinement.

\section{Solution procedure}\label{sec:procedure}

The full numerical framework integrates the FBIM for fluid-structure interactions and the OMM for sharp-interface advection-diffusion problems.
Given all values at time $t$, the solutions at time $t + \Delta t$ are calculated by following procedures:

(1) Advance particles using Eq. (\ref{eq:X});

(2) Refill solute concentrations at newly appered mesh points using Eqs. (\ref{eq:adfbi}) and (\ref{eq:adpbi});

(3) Solve advection-diffusion problems Eqs. (\ref{eq:adf}) and (\ref{eq:adp}) in each subdomain with global and interface boundary conditions, and update solute concentrations on particles;

(4) Solve for surface force $\boldsymbol{F}$, desired velocity $\boldsymbol{V}$, translational velocity $\boldsymbol{U}_p$, and rotational velocity $\boldsymbol{\mit\Omega}_p$.

(5) Solve the long range problem Eq. (\ref{eq:long}) with force density $\boldsymbol{f}^l$ and the short range problem Eq. (\ref{eq:short}) using Eq. (\ref{eq:short_range_integral_element}), and update flow field and velocities on boundaries (both particles and fixed boundaries).

\section{Validations}
\label{sec:validation}

In this section, the FBIM and the OMM are validated seperately by several numerical simulations.
Their convergence behaviors are also presented.
Here we define the relative error of any parameter $C$ as 
\begin{equation}
\varepsilon_{C} = \frac{|C_{(num)} - C_{(ref)}|}{|C_{(ref)}|},
\end{equation}
where $C_{(num)}$ and $C_{(ref)}$ are the numerical and reference solutions, respectively.

\subsection{Parabolic flow with circular boundary}
We first validate the FBIM by considering a simple parabolic flow.
The computational domain is a square of side length $L = 6.4$.
Periodic boundary conditions are employed in $x$ and $y$ directions.
A circular boundary $\mit\Gamma$ of radius 2.5 is located at coordinates $(3.0, 3.0)$, and the boundary condition reads
\begin{equation}
u_x|_{\mit\Gamma} = y^2, u_y|_{\mit\Gamma} = 0.
\end{equation}
The surface force $\boldsymbol{F}$ on the boundary is calculated by Eq. (\ref{eq:feedback_force}) with the desired velocity $\boldsymbol{V} = (y^2, 0)$.
The velocity due to this surface force is computed at coordinates $(3.2, 3.2)$ at $t = 20$.

\begin{figure}[hbt!]
\centering
	\includegraphics[width=\textwidth]{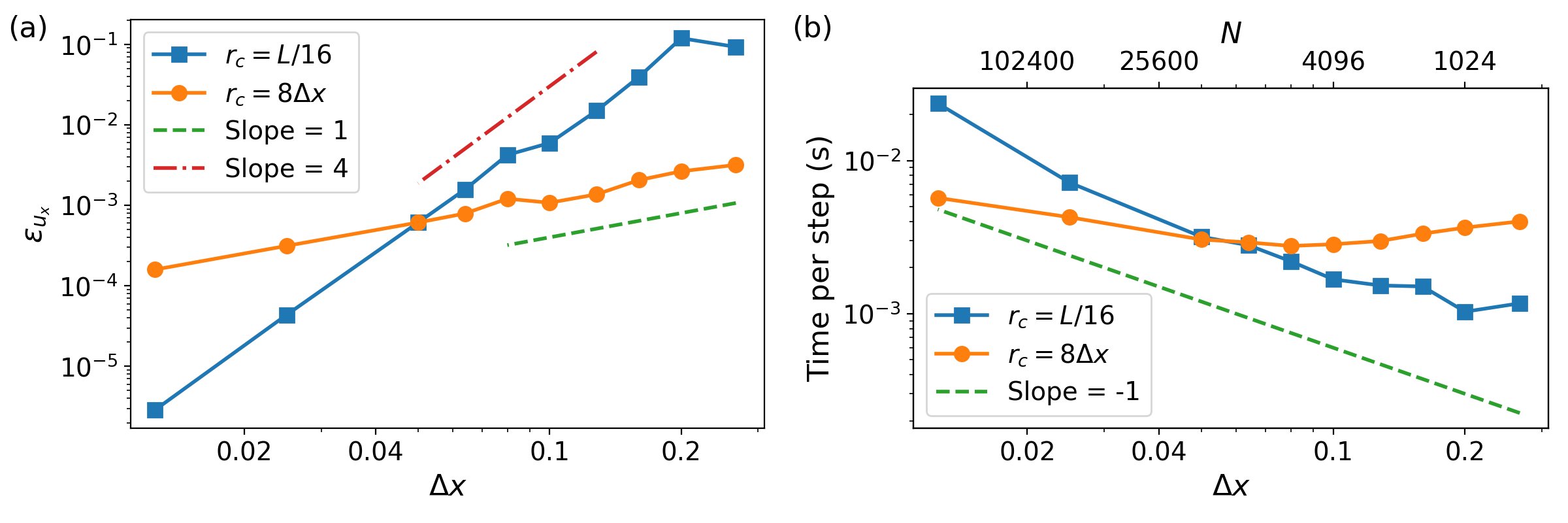}
	\caption{Validation and convergence study of the FBIM.
	(a) Convergence of velocity $u_x$ with respect to mesh size $\Delta x$ at coordinates $(3.2, 3.2)$.
	(b) {Computational time per step with respect to mesh size $\Delta x$ and number of Cartesian mesh points $N$.
	Here $N$ scales with $\Delta x^{-2}$.}
%   {\color{red}Number of time steps is also plotted as red triangles.}
	}
	\label{fig:FBIM_circle}
\end{figure}

Fig. \ref{fig:FBIM_circle} shows the relative error of $u_x$ and the computational time by varying the number of mesh points $N_x$ and $N_y$ between 24 to 512.
With fixed cutoff distance $r_c = L/16$, we observe a 4th order convergence, and {the computational time per step scales as $1 / \Delta x$.}
Here $N = N_x \times N_y$.
When setting $r_c = 8\Delta x$, the cutoff distance decreases linearly with the mesh size.
On the one hand, decreasing $r_c$ leads to a lower order convergence (1st order); on the other hand, {the computational time per step is hardly affected.
This indicates that most of the computational time is spent on the short-range problem depending on $r_c$.}
%{\color{red}Moreover, the simulation time scales as the number of time steps (see the red triangles in Fig. \ref{fig:FBIM_circle}b).}
Taking both the accuracy and efficiency into consideration, the cutoff distance is set as $r_c = 8\Delta x$ in following simulations.

\subsection{Single phoretic particle with finite system size}
\label{sec:OMM_validation}
\begin{figure}[hbt!]
\centering
	\includegraphics[width=0.3\textwidth]{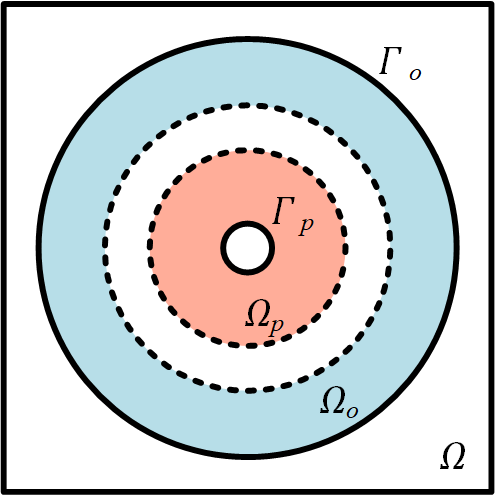}
	\caption{Schematic of the computational domain for single particle dynamics.
	The domain $\mit\Omega$ is decomposed into three overlapping subdomains: the fluid subdomain $\mit\Omega_f$ which coincides with $\mit\Omega$, the particle subdomain $\mit\Omega_p$, and the outer subdomain $\mit\Omega_o$.
	The solid lines represent the global boundaries, and the dashed lines represent the interface boundaries.}
	\label{fig:validation_schematic}
\end{figure}

For the validation of the OMM, we consider a phoretic particle with limited system size.
This problem has been studied analytically and numerically in Ref. \cite{Farutin22}.
As shown in Fig. \ref{fig:validation_schematic}, the particle boundary $\mit\Gamma_p$ and the outer boundary $\mit\Gamma_o$ are concentric circles of radius 1 and $R = 3.25$, respectively.
Here $R$ is the system size.
The governing equations are Eqs. (\ref{eq:stokes}) and (\ref{eq:ad}) with 0 consumption/production rate ($\beta = 0$).
The boundary conditions are summarized as follows \cite{Hu19}:
\begin{equation}\label{eq:Hu_u}
\boldsymbol{u}(\boldsymbol{x}, t)|_{\mit\Gamma_p} = M \nabla_s c(\boldsymbol{x}, t)|_{\mit\Gamma_p} + \boldsymbol{U}_p(t)% + \boldsymbol{\mit\Omega}(t) \times (\boldsymbol{x} - \boldsymbol{X}_p),
%\boldsymbol{u}(\boldsymbol{x}, t)|_{r = \infty} = \boldsymbol{0},
\end{equation}
\begin{equation}\label{eq:Hu}
\boldsymbol{n} \cdot \nabla c(\boldsymbol{x}, t)|_{\mit\Gamma_p} = - A,
c(\boldsymbol{x}, t)|_{\mit\Gamma_o} = 0.
\end{equation}
Moreover, the velocity attenuates to 0 in the far field.
The flow field is calculated analytically by stream function \cite{Sondak16,Hu19}.
The solute concentration is computed by the OMM in a square domain $\mit\Omega$.
As shown in Fig. \ref{fig:validation_schematic}, the computational domain $\mit\Omega$ is decomposed into three overlapping subdomains.
The fluid subdomain $\mit\Omega_f$ coincides with $\mit\Omega$, and periodic boundary conditions are used in $x$ and $y$ directions.
The particle subdomain $\mit\Omega_p$ is an annular domain comoving with the particle (see the red ring in Fig. \ref{fig:validation_schematic}).
Another annular subdomain (the outer subdomain $\mit\Omega_o$, as shown by the blue ring in Fig. \ref{fig:validation_schematic}) is attached to the outer boundary $\mit\Gamma_o$ which also comoves with the particle.
For the two annular subdomains, the global boundaries are represented by solid lines, and the boundary conditions follow Eq. (\ref{eq:Hu}), while the interface boundaries are labeled by dashed lines (see Fig. \ref{fig:validation_schematic}).

\begin{figure}[hbt!]
\centering
	\includegraphics[width=\textwidth]{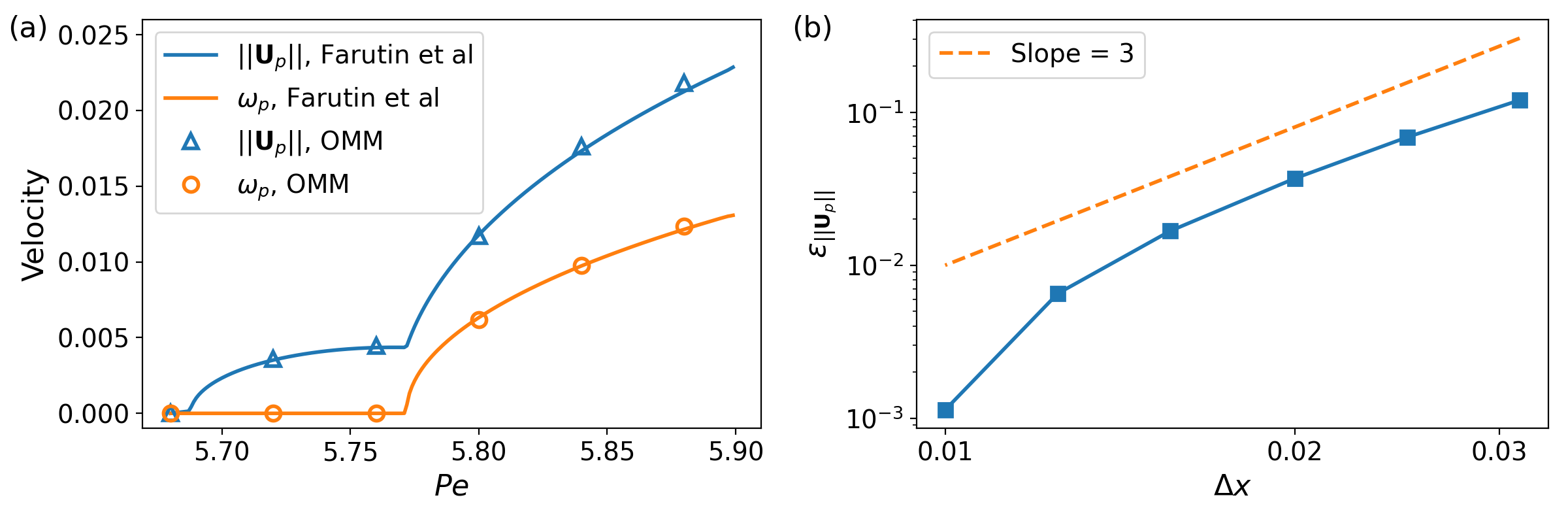}
	\caption{Comparison between the analytical prediction and numerical simulation using the OMM.
	(a) Phoretic velocity $||\boldsymbol{U}_p(t)||$ and angular velocity $\omega_p$ in stationary, straight and circular phases.
	The mesh size is $\Delta x = 1/64$.
	(b) Convergence of phoretic velocity $||\boldsymbol{U}_p||$ with respect to mesh size $\Delta x$ at $Pe = 5.72$.
	The system size is set to $R = 3.25$.
	The analytical prediction is obtained by Ref. \cite{Farutin22}.}
	\label{fig:OMM_Farutin_JFM}
\end{figure}

The particle velocity $||\boldsymbol{U}_p||$ and angular velocity $\omega_p$ obtained by the OMM are compared with the analytical results from Ref. \cite{Farutin22}, as shown in Fig. \ref{fig:OMM_Farutin_JFM}a.
Here a non-zero angular velocity indicates circular motion.
Good agreement between the two results is observed when setting the mesh size $\Delta x = 1/64$.
Convergence of the phoretic velocity $||\boldsymbol{U}_p||$ with respect to $\Delta x$ is demonstrated in Fig. \ref{fig:OMM_Farutin_JFM}b, as the P\'eclet number is $Pe = 5.72$.
To calculate the relative error $\epsilon_{||\boldsymbol{U}_p||}$, the analytical result from Ref. \cite{Farutin22} is used as a reference.
As $\Delta x$ varies between $1/100$ to $1/32$, a 3rd order accuracy is reached, thereby validating our implementation of the OMM.

\section{Applications}
\label{sec:applications}
In this section, we combine the FBIM and the OMM to explore new simulations for single particle and many particles dynamics.
First we study particle swimming in a nearly infinite domain with periodic boundary conditions.
Then particle dynamics in a straight channel is investigated.
Finally we show examples of multiple particles in a rectangular domain with periodic boundary conditions to reveal some interesting phenomena.
For all simulations, the comsumption rate is set to $\beta = 0.01$.

\subsection{Single particle in a nearly infinite domain}
We first consider a similar case as presented in section \ref{sec:OMM_validation}, but with larger system size $R = 200$, which is large enough to mimic an infinite domain \cite{Hu19}.
The computational domain is a square and its size is $L = 409.6$.
The flow field is computed on a Cartesian mesh containing 8192 grid points in each direction, corresponding to a mesh size of $\Delta x = 0.05$.
The particle surface is discretized into $N_e = 128$ elements.
The subdomain sizes are set as $r_p = 1.6$ for the two annular subdomains $\mit\Omega_p$ and $\mit\Omega_f$.
The radial and tangential meshes contain 32 and 128 grid points for $\mit\Omega_p$, and 32 and 2560 grid points for $\mit\Omega_o$, respectively.

\begin{figure}[hbt!]
\centering
	\includegraphics[width=0.8\textwidth]{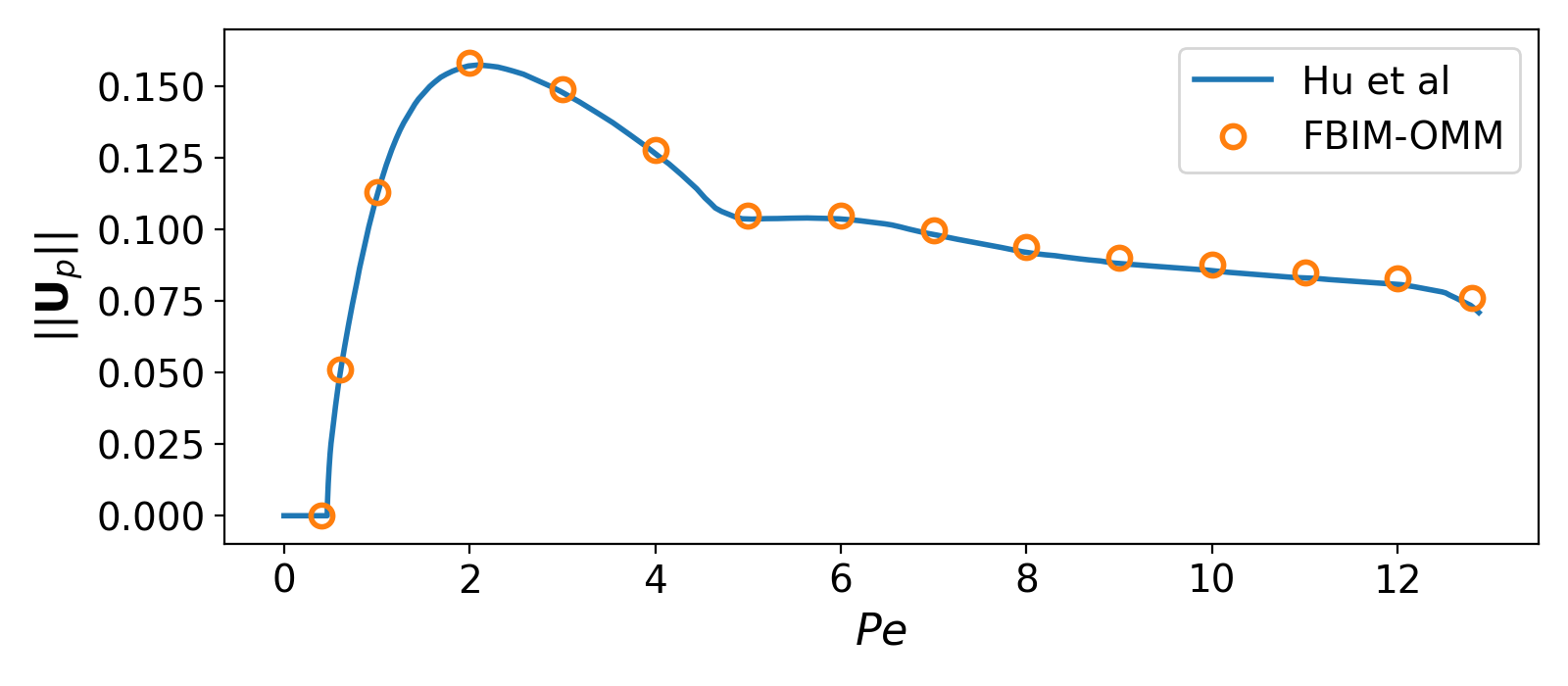}
	\caption{Phoretic velocity $||\boldsymbol{U}_p||$ as a function of $Pe$ at $R = 200$.
	The solid line represents the data obtained by Ref. \cite{Hu19}.}
	\label{fig:Umag_Hu_PRL}
\end{figure}

\begin{figure}[hbt!]
\centering
	\includegraphics[width=1.0\textwidth]{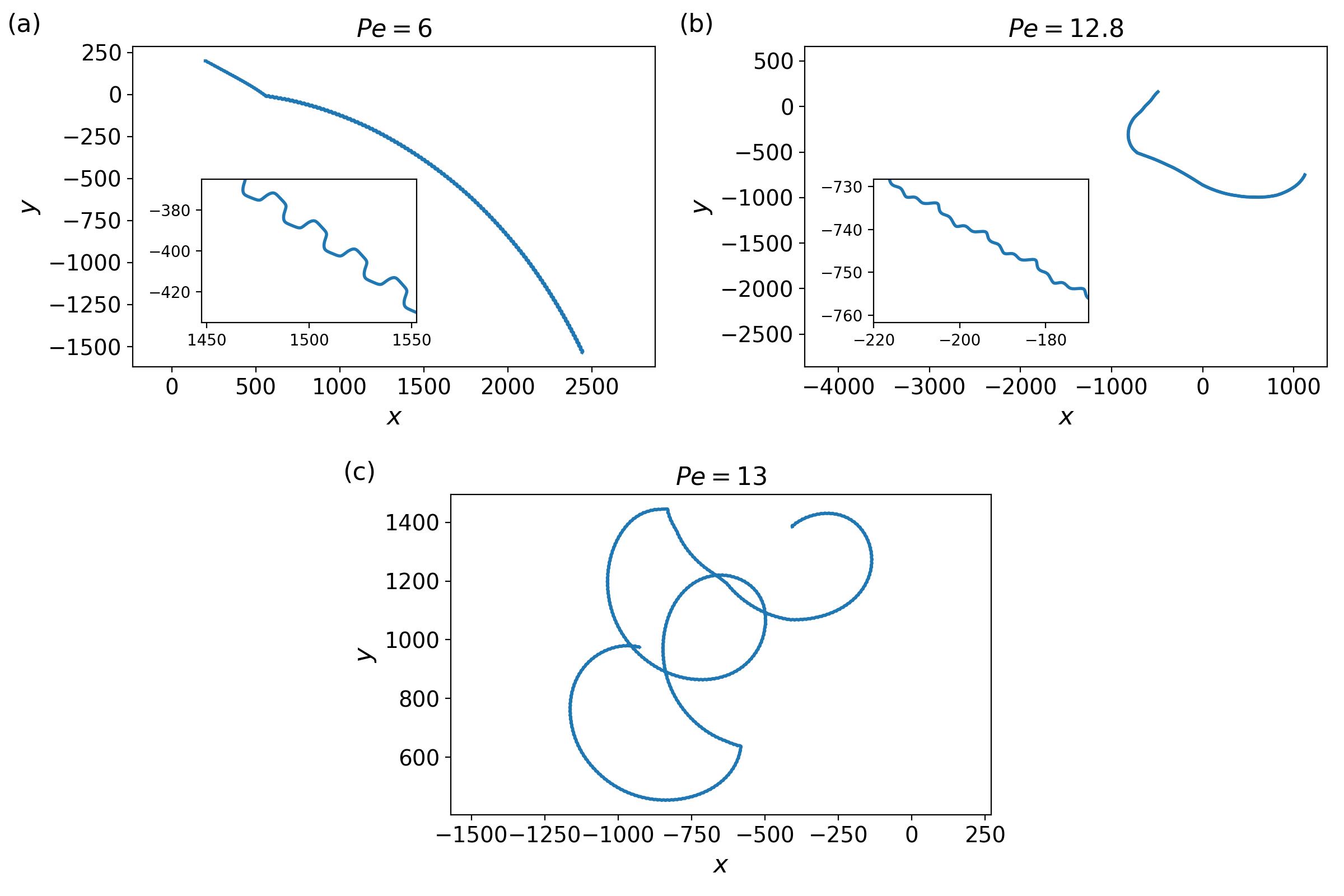}
	\caption{Different trajectories obtained by the FBIM-OMM.}
	\label{fig:Hu_trajectory}
\end{figure}

The phoretic velocity $||\boldsymbol{U}_p||$ computed by the FBIM-OMM is in good agreement with the published data \cite{Hu19} as demonstrated in Fig. \ref{fig:Umag_Hu_PRL}.
We further plot particle trajectories for $Pe = 6, 12.8, 13$ in Fig. \ref{fig:Hu_trajectory}.
At short enough time scale, the trajectories obtained by the FBIM-OMM are similar to those in Ref. \cite{Hu19}, as shown in the insets of Fig. \ref{fig:Hu_trajectory}.
However, at larger time scales, the particle does not move along a straight line at $Pe = 6$ (Fig. \ref{fig:Hu_trajectory}a), or along a circle at $Pe = 12.8$ (Fig. \ref{fig:Hu_trajectory}b), as reported in \cite{Hu19}.
{Moreover, the chaotic motion at $Pe = 13$ is different from that in Ref. \cite{Hu19} (see Fig. \ref{fig:Hu_trajectory}c).}
These deviations result from the finite computational domain size $L$ and the periodic boundary conditions.
In Ref. \cite{Hu19}, the flow velocity attenuates to 0 in the far field.
In contrast, for the present work, the flow in the computational domain is affected by its images due to periodic boundary conditions.
Although the domain size $L$ is 2 orders of magnitude larger than the particle radius, this effect is still non-negligible.
For this reason, the particle trajectories are different in long scale.

\subsection{Single particle in a straight channel}
\begin{figure}[hbt!]
\centering
	\includegraphics[width=\textwidth]{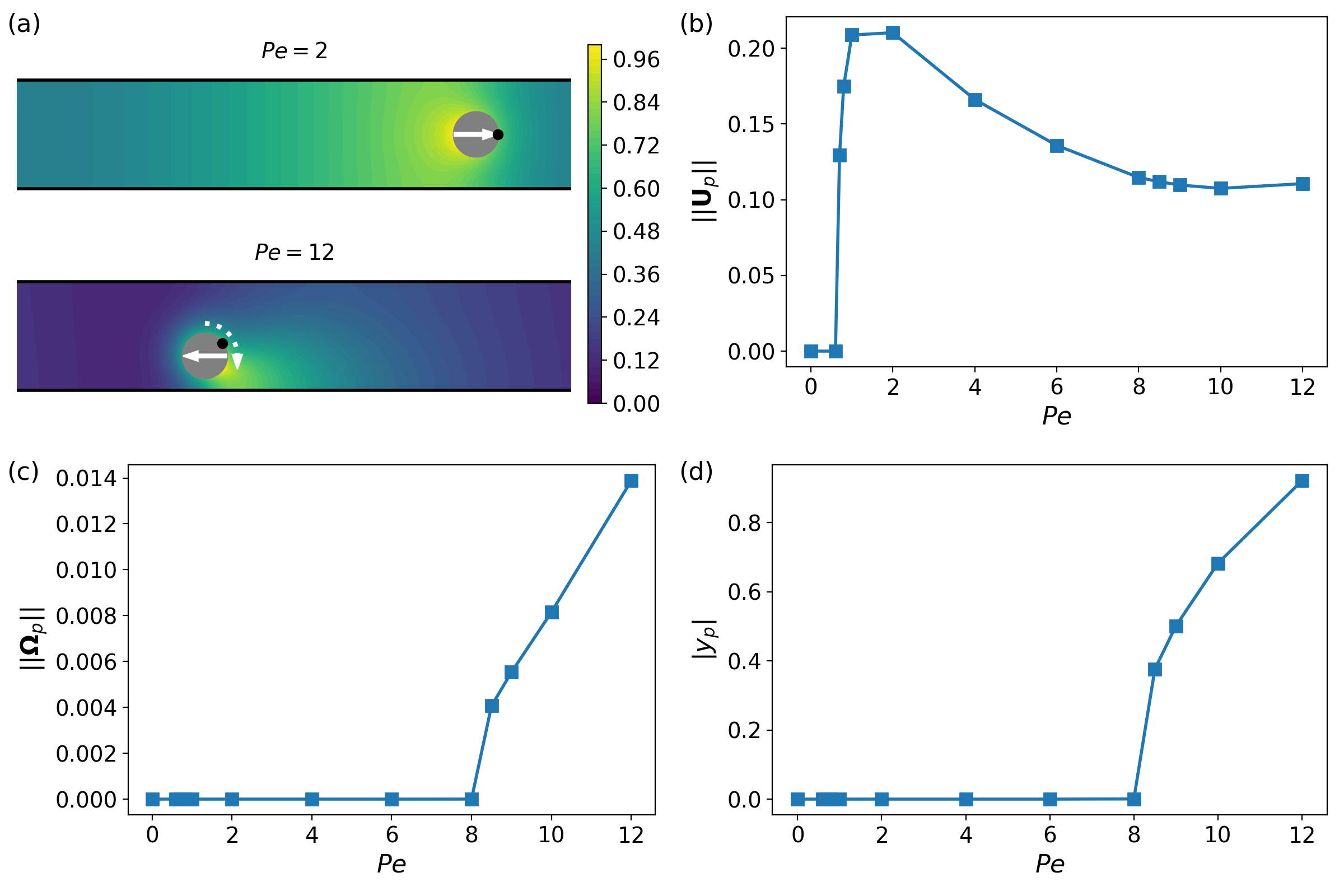}
	\caption{Single particle swimming in a straight channel.
	(a) Steady-state relative solute concentration at $Pe = 2$ and 12, where the arrows indicate translational and rotational directions.
	(b) Phoretic velocity $||\boldsymbol{U}_p||$, (c) rotational velocity $||\boldsymbol{\mit\Omega}_p||$, and (d) lateral position $|y_p|$ at steady state.
	%The unconfined solution ($Cn = 0$) is obtained from Ref. \cite{Hu19}.
	}
	\label{fig:straight_channel}
\end{figure}

In this subsection, we consider a phoretic particle immersed in a straight channel of length $L = 25.6$ and width $W = 5$, as shown in Fig. \ref{fig:straight_channel}a.
Periodic boundary conditions are used at the left and right boundaries for both the flow and solute fields.
At other boundaries, the boundary conditions follow Eqs. (\ref{eq:ub}) and (\ref{eq:cb}).
Other simulation parameters are: mesh size $\Delta x = 0.05$, particle surface element number $N_e = 128$, and particle subdomain size $r_p = 0.8$.
A convergence study with respect to $\Delta x$ and $r_p$ will be presented later in this subsection.

As shown in Fig. \ref{fig:straight_channel}a, the particle swimms in a straight line (indicated by the white solid arrows), and its velocity is shown in Fig. \ref{fig:straight_channel}b.
The phoretic velocity $||\boldsymbol{U}_p||$ shows a transition from stationary state to straight motion around $Pe \approx 0.6$, which corresponds to a pitchfork bifurcation.
Moreover, a symmetry breaking is found around $Pe \approx 8$, so that the particle stays at an off-centered position and rotates around its center (see Fig. \ref{fig:straight_channel}c and d).
Here $y_p$ is the particle lateral position from the channel center.
Indeed, when the particle locates in the center of the channel, the hydrodynamic interactions of the two fixed boundaries are equal, and no rotational motion occurs.
As the particle gets closer to one solid boundary than the other, it rotates clockwise (counter-clockwise) if the closest boundary is on the left (right) in the frame moving with the particle (see the white dashed arrow in Fig. \ref{fig:straight_channel}a).

\begin{figure}[hbt!]
\centering
	\includegraphics[width=\textwidth]{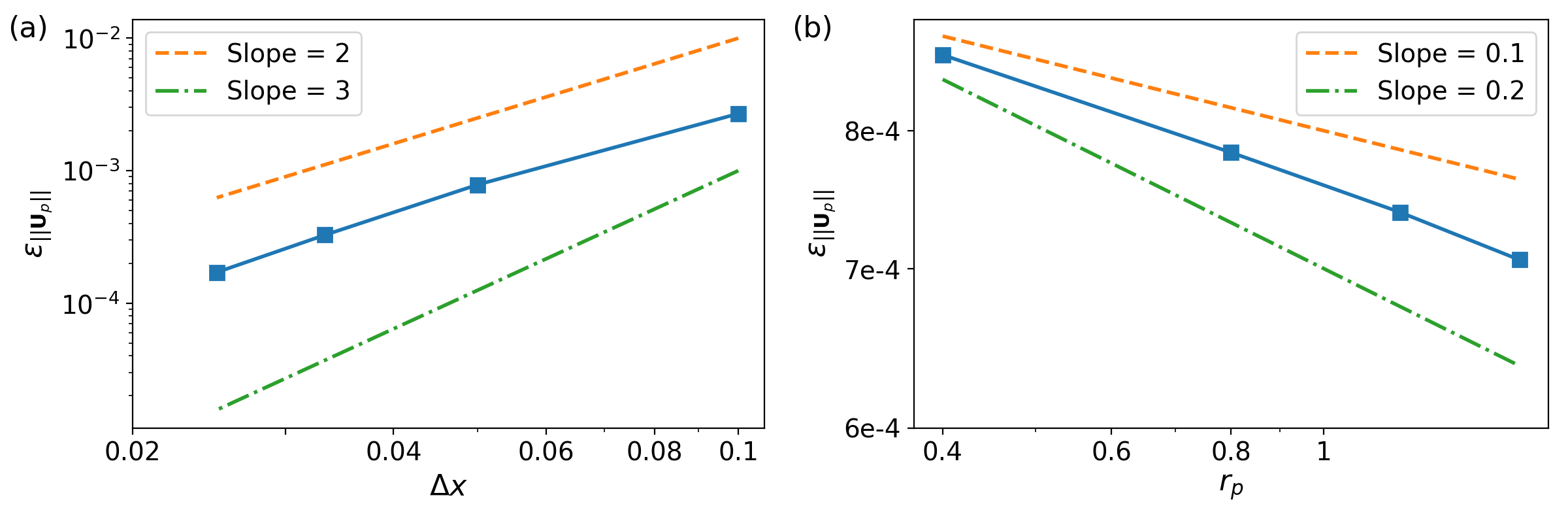}
	\caption{Convergence of phoretic velocity $||\boldsymbol{U}_p||$ with respect to (a) mesh size $\Delta x$ and (b) particle subdomain size $r_p$.
	The P\'eclet number is fixed at $Pe = 2$.}
	\label{fig:straight_Umag}
\end{figure}

The convergence test is conducted at $Pe = 2$.
The relative error of phoretic velocity $\epsilon_{||\boldsymbol{U}_p||}$ is computed, and the numerical solution for $\Delta x = 0.0125$, $N_e = 512$, and $r_p = 2.0$ is used as reference solution.
First $r_p = 0.8$ is fixed, while $\Delta x$ is varied between 0.025 to 0.1 and $N_e$ scales as $1/\Delta x$.
As shown in Fig. \ref{fig:straight_Umag}a, a second order convergence is obtained, in agreement with the nominal accuracy of the FBIM-OMM.
Then we fix the mesh size at $\Delta x = 0.05$ and the particle surface element number at $N_e = 128$, while $r_p$ is varied between 0.4 to 1.6.
The error is slightly reduced with the increase of $r_p$, as demonstrated in Fig. \ref{fig:straight_Umag}b.
In fact, larger overlap sizes exhibit better stability, while the accuracy is determined by the interpolation scheme at interface boundaries \cite{Chesshire90,Peet12}.

\subsection{Collective behaviors of multiple particles}
\begin{figure}[hbt!]
\centering
	\includegraphics[width=\textwidth]{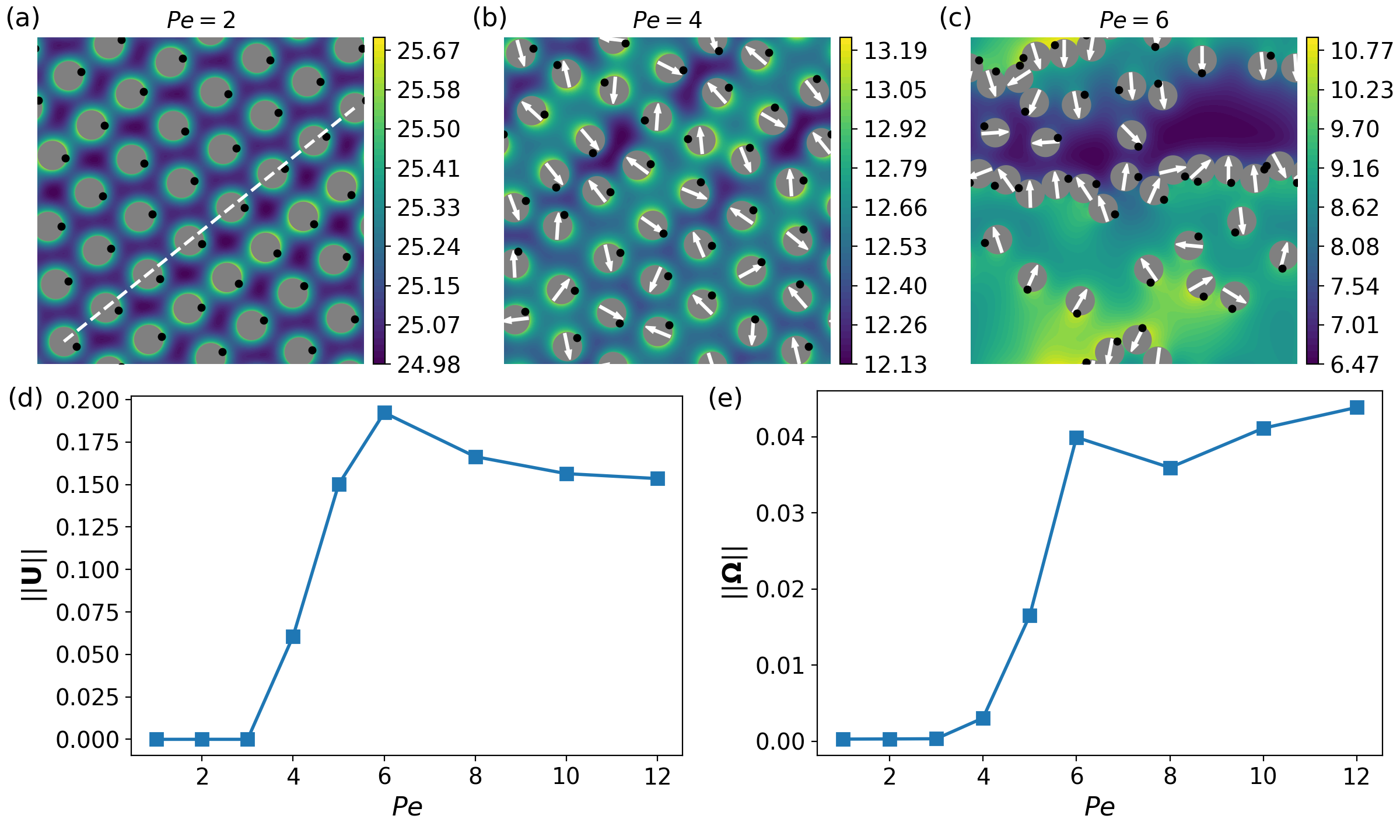}
	\caption{Collective behaviors of multiple particles.
	(a--c) Snapshots of particles and relative solute concentration at $Pe = 2$, 4, and 6.
	The white dashed line in (a) is a guide for the eyes.
	The arrows in (b) and (c) indicate translational velocity directions.
	(d) Phoretic velocity $||\boldsymbol{U}||$ and (e) rotational velocity $||\boldsymbol{\mit\Omega}||$ averaged over all particles.}
	\label{fig:cavity_Np40}
\end{figure}

In this subsection, a suspension of particles in a square domain is simulated with the proposed method.
The side length of the domain is set as $L = 25$, and 40 phoretic particles are distributed in the domain, as shown in Fig. \ref{fig:cavity_Np40}a.
The corresponding area concentration is about $20\%$.
Periodic boundary conditions are used in both directions.
According to the convergence study in the previous subsection, the other simulation parameters are chosen: mesh size $\Delta x = 0.05$, particle surface element number $N_e = 128$, and particle subdomain size $r_p = 0.8$.

At $Pe \leq 2$, the particles form a crystal-like structure at steady state, as shown in Fig. \ref{fig:cavity_Np40}a.
Although the particles stay at equilibrium positions (i.e. $||\boldsymbol{U}|| = 0$, see Fig. \ref{fig:cavity_Np40}c), there is flow among them, since the surface concentration on each particle is not homogeneous.
{Moreover, some particles rotate around their centers due to hydrodynamic interactions.
As a result, non-zero rotational velocity $||\boldsymbol{\mit\Omega}||$ is observed in Fig. \ref{fig:cavity_Np40}d.}
It should be noted that $||\boldsymbol{U}||$ and $||\boldsymbol{\mit\Omega}||$ are average values over all particles.
The crystal-like structure loses its stability for $Pe > 3$ in favor of a {liquid-like regime}, as shown in Fig. \ref{fig:cavity_Np40}b.
{Particles can swim in a short range, caged by other surrounding particles.}
For such a disordered (no noise is introduced) solution, the translational and rotational velocities are defined as their time-averaged values:
\begin{equation}
||\boldsymbol{U}|| = \frac{1}{T} \int_0^T ||\boldsymbol{U}(t)|| dt,
||\boldsymbol{\mit\Omega}|| = \frac{1}{T} \int_0^T ||\boldsymbol{\mit\Omega}(t)|| dt,
\end{equation}
which is measured over a time interval $T$.
{Increasing further $Pe$ leads to a more chaotic regime ($Pe \geq 5$), as shown in Fig. \ref{fig:cavity_Np40}c.
In this gas-like regime, both translation and orientation orders are lost, and particles can travel freely for longer distance.
Moreover, particles form traveling lines for $Pe \geq 6$ (see Fig. \ref{fig:cavity_Np40}c).
Similar behaviors have been observed experimentally \cite{Thutupalli18}.}
With the increase of $Pe$, both $||\boldsymbol{U}||$ and $||\boldsymbol{\mit\Omega}||$ show nonmonotonic behaviors, as shown in Fig. \ref{fig:cavity_Np40}d and e.
These results lay the foundation for investigations of collective motions of phoretic particles.
We leave detailed explanation and systemic studies in future works.

\section{Conclusion}
\label{conclusion}
A full numerical framework for phoretic particles was presented.
This framework consists of a FBIM for fluid-structure interactions in Stokes flow, and an OMM for advection-diffusion problems with moving boundaries.
The FBIM decomposes the Stokes equations into short-range and long-range parts, so that the former is computed by boundary integral equation within a cut-off distance, while the latter is solved by a Fourier spectral method accelerated by FFT.
To resolve moving boundaries in advection-diffusion problems, the OMM divides the computational domain into several overlapping subdomains, which are allowed to move independently.
The FBIM and the OMM were then validated separately, showing relatively high order accuracy.
We further applied this framework to more general problems, including single particle in nearly infinite domain, single particle in a straight channel, and collective behaviors of multiple particles.
This framework can be further extended to non-circular rigid particles and even deformable particles, which will be our future research.

%\section*{Declaration of competing interests}
%The authors declare that they have no known competing financial interests or personal relationships that could have appeared to influence the work reported in this paper.

\section*{Acknowledgements}
This work was supported by CNES (Centre National d'Etudes Spatiales) and the French-German university program ``Living Fluids'' (Grant CFDA-Q1-14). 
The simulations were performed on the Cactus cluster of the CIMENT infrastructure, which is supported by the Rh\^one-Alpes region (GRANT CPER07\_13 CIRA).


\appendix

\section{Long-range density}\label{sec:appendix_a}
A natural choice of $D^l(r)$ is the polynomial, of which the high order derivatives are continuous.
In practice, we chose a 6th-order polynomial $D^l(r) = \sum_{i = 0}^6 a_i r^i$ and set 
\begin{equation}\label{eq:Dl_dr}
\left. \frac{\partial^k D^l}{\partial r^k} \right|_{r = r_c} = 0, \left. \frac{\partial D^l}{\partial r} \right|_{r = 0} = 0,
\end{equation}
where $k = 0, 1, 2, 3$.

The integral in Eq. (\ref{eq:long_range_green}) can be writen in polar coordinates, and the following equation is obtained as
\begin{equation}
G_{ij}(\boldsymbol{x}, \boldsymbol{x}^v) = \int_0^{r_c} D^l(r') r' \left[ \int_{0}^{2\pi} G_{ij}(\boldsymbol{r}, \boldsymbol{r}') d\theta \right] dr',
\end{equation}
where $\boldsymbol{r}' = \boldsymbol{x}' - \boldsymbol{x}^v$ and $\theta$ is the angle between $\boldsymbol{r}$ and $\boldsymbol{r}'$.
Substitute Eq. (\ref{eq:green}) into the equation above and integrate over $\theta$, we obtain the following equation
\begin{equation}\label{eq:long_range_constraint}
G_{ij}(\boldsymbol{x}, \boldsymbol{x}^v) = 2\pi G_{ij}(\boldsymbol{x}, \boldsymbol{x}^v) \int_0^{r_c} D^l(r') r' dr' + \pi \frac{r_i^* r_j^* - r_i r_j}{r^4} \int_0^{r_c} D^l(r') r'^3 dr'.
\end{equation}
Here $\boldsymbol{r}^*$ is the vector perpendicular to $\boldsymbol{r}$.
To satisfy (\ref{eq:long_range_constraint}), we set 
\begin{equation}\label{eq:Dl_rc}
\int_0^{r_c} D^l(r') r' dr' = \frac{1}{2\pi}, \int_0^{r_c} D^l(r') r'^3 dr' = 0.
\end{equation}

By solving the system of (\ref{eq:Dl_dr}) and (\ref{eq:Dl_rc}), we obtain the polynomial coefficients, and the long-range density is given as 
\begin{equation}
D^l(r) = \left\{
\begin{aligned}
&0, & r \geq r_c \\
&\frac{56}{3 \pi r_c^2} \left[-\frac{25}{2}(\frac{r}{r_c})^6 + 54(\frac{r}{r_c})^5 - 90(\frac{r}{r_c})^4 + 70(\frac{r}{r_c})^3 - \frac{45}{2}(\frac{r}{r_c})^2 + 1 \right], & r < r_c
\end{aligned}
\right.
.
\end{equation}

\section{Short-range Green's function}\label{sec:appendix_b}
Rewrite the integral in Eq. (\ref{eq:us}) in polar coordinate and substitute Eqs. (\ref{eq:Dl}) and (\ref{eq:Ds}), the short-range boundary integral equation becomes:
\begin{equation}
\begin{aligned}
u_i^s(\boldsymbol{x}) &= \frac{1}{4 \pi} g_j^v G_{ij}(\boldsymbol{x}, \boldsymbol{x}^v) - \frac{1}{4 \pi} g_j^v \int_0^{r_c} D^l(r') r' \left[ \int_{0}^{2\pi} G_{ij}(\boldsymbol{r}, \boldsymbol{r}') d\theta \right] dr' \\
&= \frac{1}{4 \pi} g_j^v G_{ij}(\boldsymbol{x}, \boldsymbol{x}^v) - A_1 g_j^v G_{ij}(\boldsymbol{x}, \boldsymbol{x}^v) - A_2 g_j^v (r_i^* r_j^* - r_i r_j) - A_3 g_j^v \delta_{ij} - A_4 g_j^v \delta_{ij}.
\end{aligned}
\end{equation}
where
\begin{equation}
\begin{split}
A_1 &= \frac{1}{2} \int_0^r D^l(r') r' dr' \\
    &= \frac{28}{3 \pi} \left[ - \frac{25}{16} (\frac{r}{r_c})^8 + \frac{54}{7} (\frac{r}{r_c})^7 - 15 (\frac{r}{r_c})^6 + 14 (\frac{r}{r_c})^5 - \frac{45}{8} (\frac{r}{r_c})^4 + \frac{1}{2} (\frac{r}{r_c})^2 \right], \\
A_2 &= \frac{1}{4 r^4} \int_0^r D^l(r') r'^3 dr' \\
    &= \frac{14}{3 \pi r_c^2} \left[ - \frac{5}{4} (\frac{r}{r_c})^6 + 6 (\frac{r}{r_c})^5 - \frac{45}{4} (\frac{r}{r_c})^4 + 10 (\frac{r}{r_c})^3 - \frac{15}{4} (\frac{r}{r_c})^2 + \frac{1}{4} \right], \\ 
A_3 &= - \frac{1}{2} \int_r^{r_c} D^l(r') r' \ln(r') dr'\\
    &= - \frac{28}{3 \pi} \left[ - \frac{25}{128} (\frac{r}{r_c})^8 + \frac{54}{49} (\frac{r}{r_c})^7 - \frac{5}{2} (\frac{r}{r_c})^6 + \frac{14}{5} (\frac{r}{r_c})^5 - \frac{45}{32} (\frac{r}{r_c})^4 + \frac{1}{4} (\frac{r}{r_c})^2 - \frac{1583}{31360} \right. \\
    & + \frac{25}{16} (\frac{r}{r_c})^8 \ln(r) - \frac{54}{7} (\frac{r}{r_c})^7 \ln(r) + 15 (\frac{r}{r_c})^6 \ln(r) - 14 (\frac{r}{r_c})^5 \ln(r) + \frac{45}{8} (\frac{r}{r_c})^4 \ln(r) \\
    &\left. - \frac{1}{2} (\frac{r}{r_c})^2 \ln(r) + \frac{3}{112} \ln(r_c) \right], \\
A_4 &= \frac{1}{4} \int_r^{r_c} D^l(r') r' dr'\\
    &= \frac{14}{3 \pi} \left[ \frac{25}{16} (\frac{r}{r_c})^8 - \frac{54}{7} (\frac{r}{r_c})^7 + 15 (\frac{r}{r_c})^6 - 14 (\frac{r}{r_c})^5 + \frac{45}{8} (\frac{r}{r_c})^4 - \frac{1}{2} (\frac{r}{r_c})^2 + \frac{3}{112} \right].
\end{split}
\end{equation}
Then the short-range Green's function is obtained as
\begin{equation}
\begin{aligned}
G_{ij}^s (\boldsymbol{x}, \boldsymbol{x}^v) &= \int_{\mit\Omega_c} G_{ij}(\boldsymbol{x}, \boldsymbol{x}') D^s(|\boldsymbol{x}' - \boldsymbol{x}^v|) d\boldsymbol{x}'\\
 &= (1 - 4 \pi A_1) G_{ij} (\boldsymbol{x}, \boldsymbol{x}^v) - 4 \pi A_2 (r_i^* r_j^* - r_i r_j) - 4 \pi (A_3 + A_4) \delta_{ij}.
\end{aligned}
\end{equation}

\bibliographystyle{elsarticle-num} 
\bibliography{refs}


%\begin{thebibliography}{00}


%\bibitem{}

%\end{thebibliography}
\end{document}